\documentclass[twocolumn]{aastex631}
\usepackage{url}
\shorttitle{TU Tau B}
\shortauthors{Gray et al.}
\graphicspath{{./}{figures/}}

\begin{document}

\title{TU Tau B: The Peculiar ``Eclipse'' of a possible proto-Barium
  Giant}

\author[0000-0001-7588-0477]{Richard O. Gray}
\affiliation{Department of Physics and Astronomy, Appalachian State University
  Boone, NC 28608, USA}

\author[0000-0001-6797-887X]{Christopher J. Corbally}
\affiliation{Vatican Observatory Research Group, Steward Observatory, Tucson, AZ 85721-0065 USA}

\author[0009-0005-0817-1183]{Michael M. Briley}
\affiliation{Department of Physics and Astronomy, Appalachian State University
  Boone, NC 28608, USA}

\author[0000-0002-0622-2400]{Adam McKay}
\affiliation{Department of Physics and Astronomy, Appalachian State University
  Boone, NC 28608, USA}

\author{Forrest Sims}
\affiliation{Desert Celestial Observatory, Gilbert, AZ 85233, USA}

\author{David Boyd}
\affiliation{BAA Variable Star Section, West Challow Observatory, Wantage OX12 9TX, UK}

\author{Christophe Boussin}
\affiliation{Observatoire de l'Eridan et de la Chevelure de B\'er\'enice, 02400 \'Epaux-B\'ezu , France}

\author[0000-0001-6967-1142]{Courtney E. McGahee}
\affiliation{Department of Physics and Astronomy, Appalachian State University
  Boone, NC 28608, USA}

\author{Robert Buchheim}
\affiliation{Lost Gold Observatory, 8731 E. Lost Gold Cir, Gold Canyon, AZ
  85118, USA}

\author{Gary Walker}
\affiliation{Maria Mitchell Observatory, Minor Planet Center \#811, 4 Vestal Street, Nantucket, MA 02554 USA}

\author{David Iadevaia}
\affiliation{Mountain View Observatory, 5700 N Avenida Observatory, Tucson, AZ 85750 USA}

\author{David Cejudo Fernandez}
\affiliation{Camino de las Canteras, 42, El Berruero 28192, Spain}

\author{Damien Lemay}
\affiliation{10  Observatoire St-Anaclet, 195 Rang 4 Ouest, St-Anaclet, Quebec, G0K 1H0, Canada}

\author{Jack Martin}
\affiliation{Huggins Spectroscopic Observatory, Rayleigh, Essex, SS6 8AW, UK}

\author{Jim Grubb}
\affiliation{SAROS-1, 51000 Smith Road, Bradley, California, 93426 USA}

\author{Albert Stiewing}
\affiliation{Desert Wing Observatory, 16210 N. Desert Holly Dr., Sun City, AZ
  85351, USA}

\author{Joseph Daglen}
\affiliation{Daglen Observatory, Mayhill, New Mexico, 88339, USA}

\author{Keith Shank}
\affiliation{DEVO Observatory, Carrollton, Texas, 75007, USA}

\author{Sydney Andrews}
\affiliation{Department of Physics and Astronomy, Appalachian State University
  Boone, NC 28608, USA}

\author{Nick Barnhardt}
\affiliation{Department of Physics and Astronomy, Appalachian State University
  Boone, NC 28608, USA}

\author{Rebekah Clark}
\affiliation{Department of Physics and Astronomy, Appalachian State University
  Boone, NC 28608, USA}

\author{Hunter Corman}
\affiliation{Department of Physics and Astronomy, Appalachian State University
  Boone, NC 28608, USA}

\author{Sabina Gomes}
\affiliation{Department of Physics and Astronomy, Appalachian State University
  Boone, NC 28608, USA}

\author{Agastya Jonnalagadda}
\affiliation{Department of Physics and Astronomy, Appalachian State University
  Boone, NC 28608, USA}

\author{Theo McDaries}
\affiliation{Department of Physics and Astronomy, Appalachian State University
  Boone, NC 28608, USA}

\author{Ava Mills}
\affiliation{Department of Physics and Astronomy, Appalachian State University
  Boone, NC 28608, USA}

\author{Will Newsom}
\affiliation{Department of Physics and Astronomy, Appalachian State University
  Boone, NC 28608, USA}

\author{Andrew Slate}
\affiliation{Department of Physics and Astronomy, Appalachian State University
  Boone, NC 28608, USA}

\author{Michael Watts}
\affiliation{Department of Physics and Astronomy, Appalachian State University
  Boone, NC 28608, USA}

\begin{abstract}
TU Tau (= HD~38218 = HIP~27135) is a binary system consisting of a C-N carbon
star primary and an A-type secondary.  We report on new photometry and
spectroscopy which tracked the recent disappearance of the A-star secondary.
The dimming of the A-star was gradual and irregular, with one or more brief
brightenings,  implying the presence of nonhomogeneities in the carbon star
outflow.  We also present evidence that the A-star
is actively accreting {\it s}-process enriched material from the carbon star
and suggest that it will therefore eventually evolve into a Barium giant.  This
is an important system as well because the A-type star can serve as a probe of
the outer atmosphere of the carbon star.
\end{abstract}


\section{Introduction}
\label{sec:intro}

TU Tau is a binary system consisting of a C-N type carbon star primary and
an A-type secondary.  The carbon star primary has extremely low
flux in the blue-violet, and so flux from the A-star companion can dominate
in that part of the spectrum (see Figure 1).  \citet{sanford44} was the first
to discover the composite nature of the spectrum of TU Tau.  \citet{barnbaum96}
classified the carbon star primary as C-N4$^+$ C$_2$6, and the companion was
classified by \citet{richer72} as A2 III.  We will update these
classifications in section \ref{sec:class}.
C-N giants show enhanced abundances of the {\it s-}process elements (such
as strontium,
yttrium, zirconium, and barium), and are believed to be Asymptotic Giant
Branch (AGB) stars.  The carbon star in the TU Tau system is a semi-regular
variable (classified as SRb) with an approximate period of 190 days \citep{gcvs}\footnote{General
  Catalog of Variable Stars, June 2022 version \url{https://heasarc.gsfc.nasa.gov/W3Browse/all/gcvs.html}},
although \citet{koen02} report a period of 366 days.

\begin{figure*}
  \includegraphics[width=\textwidth]{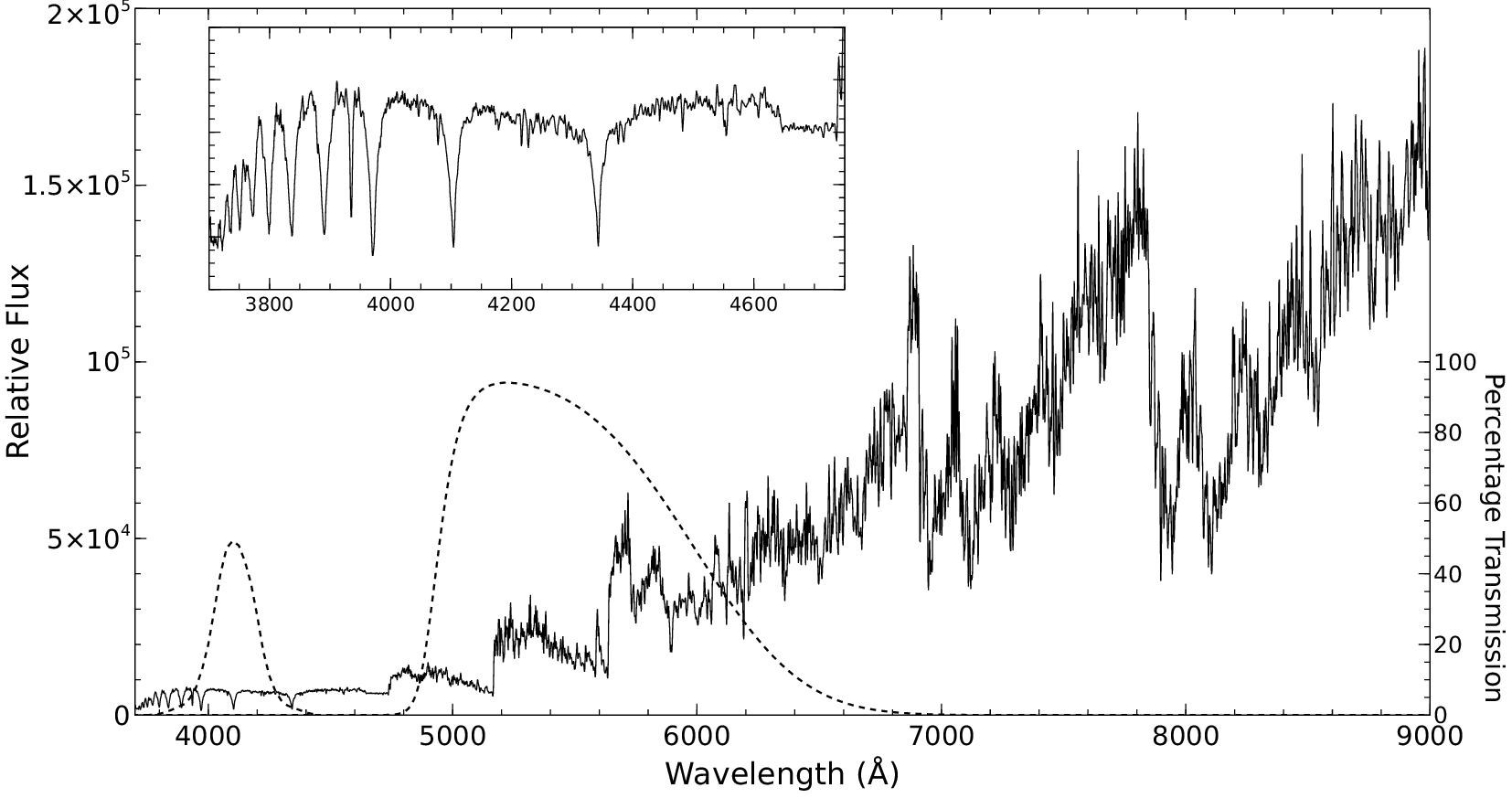}
  \caption{The LAMOST Dec 20, 2013 spectrum of TU Tau.  Flux from the carbon
    star dominates at the longer wavelengths whereas the A-star companion (also
    shown in the inset) dominates in the blue-violet.  Also included in this
    plot are the passbands for the Str\"{o}mgren-$v$ filter (dashed, centered
    near 4100\AA) and the Johnson-$V$ filter (dashed, centered near 5500\AA).
    It is clear that the flux through the Str\"{o}mgren-$v$ filter is almost
    exclusively from the A-star companion.  The flux through the Johnson-$V$
    filter is dominated by the carbon star with a small contribution from the
    A-star companion.}
  \label{fig:fig1}
\end{figure*}

\citet{mason01} attempted to resolve the binary using speckle interferometry,
but failed, implying an angular separation $\rho <= 0.056''$.
\citet{kervella19} included TU Tau
in their study of binary stars using
observed accelerations in the {\it Gaia} proper motions.  Proper motion accelerations
are indeed observed in the {\it Gaia} data for TU Tau implying detection
of the orbital motion, but the orbital characteristics, including the orbital
period, are still undetermined.  Nevertheless, \citet{kervella19} list estimates
for the masses of the two components.  For the primary, they derive a mass of
$4.05 \pm 0.2 M_\odot$, which is reasonable for an AGB star, but the derived mass
for the secondary is given as $522^{+156}_{-78} M_{\rm jup} \approx 0.5 M_\odot$ which, as we
will show in Section \ref{sec:param}, is inconsistent with the spectral
type of the secondary.   Despite the fact that the orbital period is
  currently unknown, it is possible to compute a rough lower limit.
  \citet{vanbelle13} give an angular diameter for TU Tau A $= 3.894 \pm 0.011$ mas.
  Combining that with the Gaia DR3 parallax \citep{gaia2022k}
  $= 0.8482 \pm 0.0341$ mas yields $R({\rm{TU~Tau~A}}) \approx 500R_\odot
    \approx 2.3$ AU.
    Assuming a total mass $M_A + M_B = 7M_\odot$ (see Section \ref{sec:param})
    and that the two stars are in contact, we derive
    $P_{\rm min} \approx 1.3$ yrs.

We will discuss evidence in Sections \ref{sec:class} and \ref{sec:analysis}
that the A-star
component is undergoing active accretion of material from the carbon star.
As it turns out, the interaction goes the other direction as well.  The
A-star companion affects the infrared spectrum of the
carbon star.  \citet{boersma06} have
detected unidentified infra-red (UIR) emission due to fluorescence by polycyclic
aromatic hydrocarbon (PAH) molecules in the ISO SWS mid-infrared spectrum of
TU Tau.  The UIR emission requires excitation by UV photons of the PAH
molecules, and those photons are presumably provided by the A-star
companion.  The authors also argue that the UV photons from the companion
may result in complex photochemistry in the ejecta from the carbon star.
\citet{reiter15} published Spitzer/IRAC colors for TU Tau and note that it
shows the largest IR excess for any of the semi-regular variables in their
sample.

\begin{figure*}
  \includegraphics[width=\textwidth]{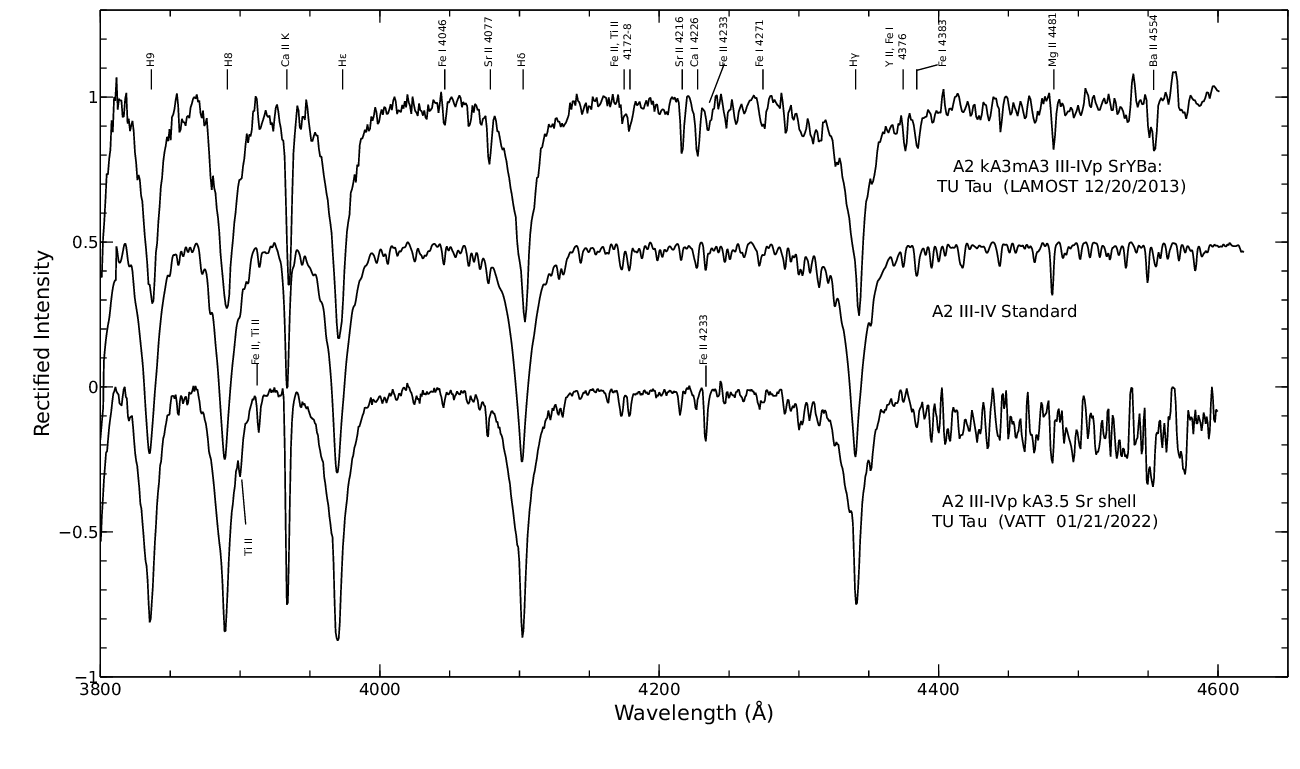}
  \caption{Top: The blue-violet portion of the LAMOST spectrum of TU Tau B
    showing the spectrum of the A-star companion in detail.  Note the strong
    lines of {\it s}-process elements, including strontium, barium and yttrium.
    Center: A standard A2 III-IV spectrum from the {\tt libr18} standards
    library for comparison.
    Bottom: A spectrum of TU Tau B obtained with the Vatican Advanced
    Technology spectrograph on 01/21/2022.  Note the strength of the Fe II
    $\lambda 4233$ line.  See Section \ref{sec:class} for more details on the
    classification of these
    spectra.  All three spectra have
    been normalized to the continuum but the bottom two spectra have
    been displaced by 0.5 and 1.0 continuum units respectively for clarity.
  }
  \label{fig:classification}
\end{figure*}

Our study of TU Tau came about through an interest in discovering the
progenitors of the barium giant stars.  Barium giants are G- and K-type giants
that show enhanced abundances of the {\it s-}process elements at an
evolutionary stage well before the {\it s-}process elements are produced by
neutron capture and dredged up to the surface.  Usually, {\it s-}process
enhancement occurs during the AGB evolutionary stage in C-N type carbon
stars.  For the barium stars, the prevailing theory for the origin of the
{\it s-}process enhancements is external contamination via mass transfer to the
current Ba star from an AGB companion that has now evolved to a white dwarf
\citep{mcclure80}.  Simple arguments based on evolutionary timescales
and the orbital characteristics of Ba giants \citep{north00} on the one hand,
and white dwarf cooling times \citep{bohm-vitense00} on the other, suggest
that most Ba giants must have been contaminated while on the main sequence.
Since  most Ba giants have masses $\ge 2.0 M_\odot$ \citep{escorza17},
this implies that on the main sequence these stars were B- or A-type.  A
search of carbon stars in the LAMOST \citep{wang96}
Data Release 7 spectrum database\footnote{\url{http://www.lamost.org/lmusers/}}
revealed a number of C-N stars with
A-type companions, including TU Tau.  Remarkably, a close inspection of the
spectrum of the A-star companion in the TU Tau system revealed an otherwise
normal A-star with strong lines of Sr~II, Ba~II, and Y~II, suggesting that
it is a prime candidate for a barium giant progenitor (see Figure
\ref{fig:classification}, top spectrum).
This prompted a
campaign of spectroscopy and photometry, eventually involving not only
the Dark Sky Observatory (Appalachian State University, North Carolina) and
the Vatican Advanced Technology Telescope spectrograph, but also a number of
amateur observatories which have made significant contributions to this
research.

\section{Observations}

\subsection{Photometry}

Robotic CCD photometry of TU Tau in four bands, Str\"{o}mgren-{\it v}
(S-{\it v}, centered at 4100\AA), and
Johnson-Cousins $B$, $V$, and $R_C$ has been carried out at Appalachian State
University's Dark Sky Observatory
(DSO, situated in the Blue Ridge Mountains of North Carolina at an elevation
of 1000 meters) using the wide-field imager in the DSO-6 dome since
October 2021.  That
imager consists
of a 300mm f/4 camera lens, and a filter wheel and CCD from Finger Lakes
Instruments.  The CCD chip is a KAF-16803.  Full frame images cover a $7^\circ
\times 7^\circ$ field on the sky.  The data are reduced with a pipeline
using functions from the python ccdproc package
\citep{matt_craig_2017_1069648}.  Differential
aperture photometry is carried out with functions from the python library
photutils \citep{larry_bradley_2023_7946442}.  Exposures (S-{\it v}: $5 \times 90$s, B: $5 \times 90$s,
V: $5 \times 45$s, R: $5 \times 10$s) were obtained each clear night TU Tau
could be observed.
The comparison and check stars were TYC 1866-2430-1 (designated on the AAVSO
chart as star ``98'')
and TYC 1866-990-1 (star ``107'').  The adopted magnitudes
for those stars on the standard system \citep{henden09} are given in Table 1.
The standard magnitudes for the S-{\it v} band have not been determined for
those
comparison stars.  However, the star HD~38334, which lies about $15'$ from
TU Tau has published Str\"{o}mgren {\it uvby} photometry, which we use in
Section \ref{sec:param} to
estimate the intrinsic S-{\it v} magnitude and by extension the absolute
$V$ magnitude of the A-star component.

\begin{table}
  \caption{Standard Johnson $B$ and $V$ magnitudes}
  \label{tab:mag}
  \begin{tabular}{lcc}
    \hline
    \hline
  filter & star 98 & star 107\\
  \hline
  $B$ & 10.897 (0.060) & 11.804 (0.069)\\
  $V$ & 9.799 (0.004) & 10.726 (0.054)\\
  \hline
  \end{tabular}
\end{table}

Table \ref{tab:BVR} lists the sources of $B$, $V$, and $R_C$ photometry utilized
for this study.

\begin{table*}
  \caption{Sources of $B$, $V$, and $R_C$ Photometry}
  \label{tab:BVR}
  \resizebox{\textwidth}{!}{%
  \begin{tabular}{lcllll}
    \hline
    \hline
    Observatory & Telescope Diam (m) & Filters & Location & Observer & Abbreviation\\
    \hline
    Dark Sky Observatory & 0.08 & S-{\it v}, $B$, $V$, $R_C$ & North Carolina & Richard Gray & DSO\\ 
    AAVSONet BSM NH2 & 0.18 & $B$, $V$, $R_C$ & New Hampshire & Robert Buchheim & BHU \\
    Sierra Remote Observatory & 0.51 & $V$ & California & Gary Walker & WGR\\
    Desert Celestial Observatory & 0.1  & $V$ & Arizona  & Forrest Sims  & SFOA\\
    Mountain View Observatory & 0.1 & $V$ & Arizona   & David Iadevaia  & IDG\\
    West Challow Observatory & 0.36 & $B$, $V$, $R_C$ & UK & David Boyd & BDG\\
    Observatorio El Gallinero & 0.41 & $B$, $V$ & Spain & David Cejudo Fernandez & CDZ\\
    Observatoire St-Anaclet &  0.36 & $B$, $V$, $R_C$ &  Quebec  &  Damien Lemay & LMA\\
    \hline
  \end{tabular}}
\end{table*}

\begin{figure}
  \includegraphics[width=\columnwidth]{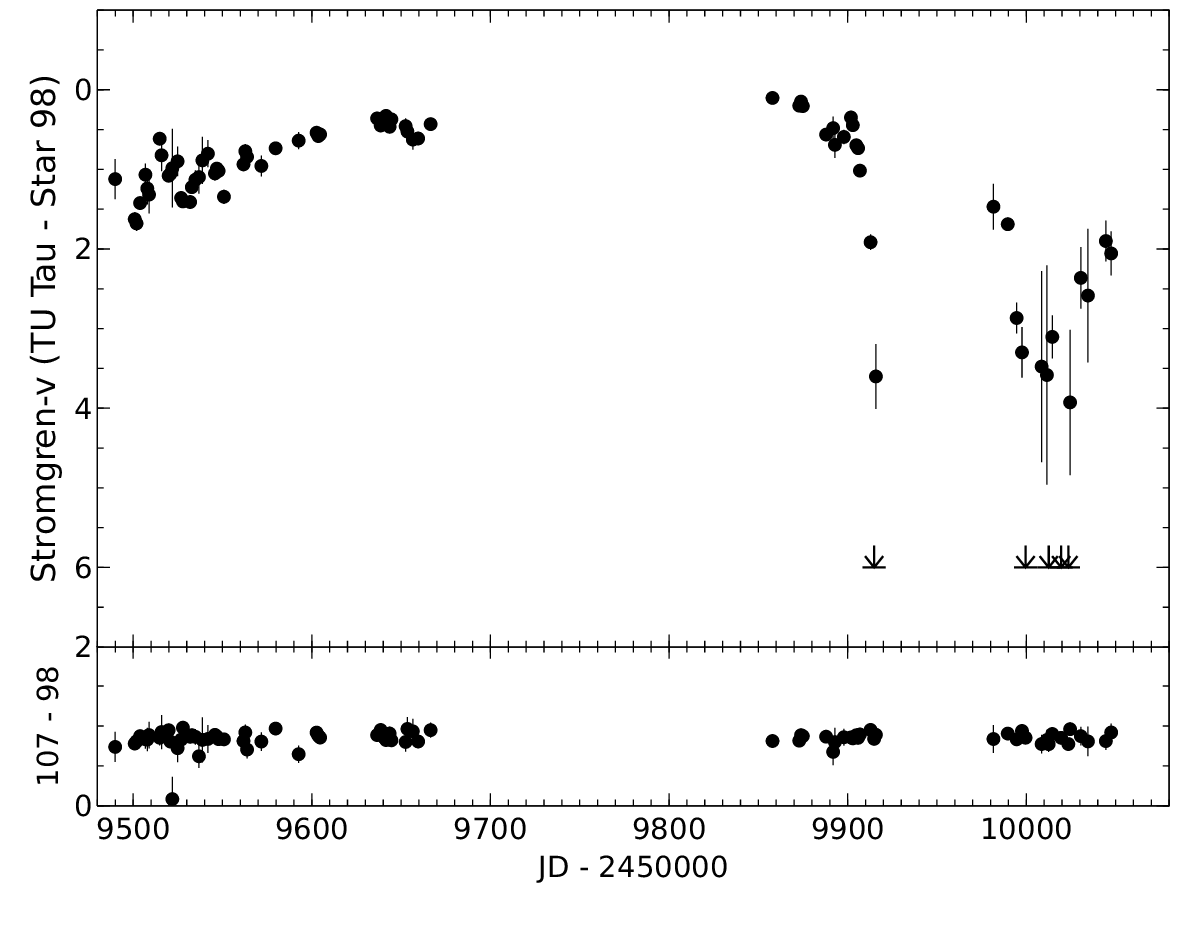}
  \caption{Str\"{o}mgren {\it v}-band photometry (centered at 4100\AA)
    obtained with
    the wide-field imager at the Dark Sky Observatory.  The top panel shows
    the photometry of TU Tau relative to the comparison star `98', whereas
    the lower panel (on the same vertical scale) shows the magnitude difference
    between the comparison star (`98') and the check star `107'.  Five points
    in the upper panel are upper limits.  TU Tau is not detected in those
    frames, even with a close visual inspection.}
  \label{fig:Sv}
\end{figure}

\begin{figure}
  \includegraphics[width=\columnwidth]{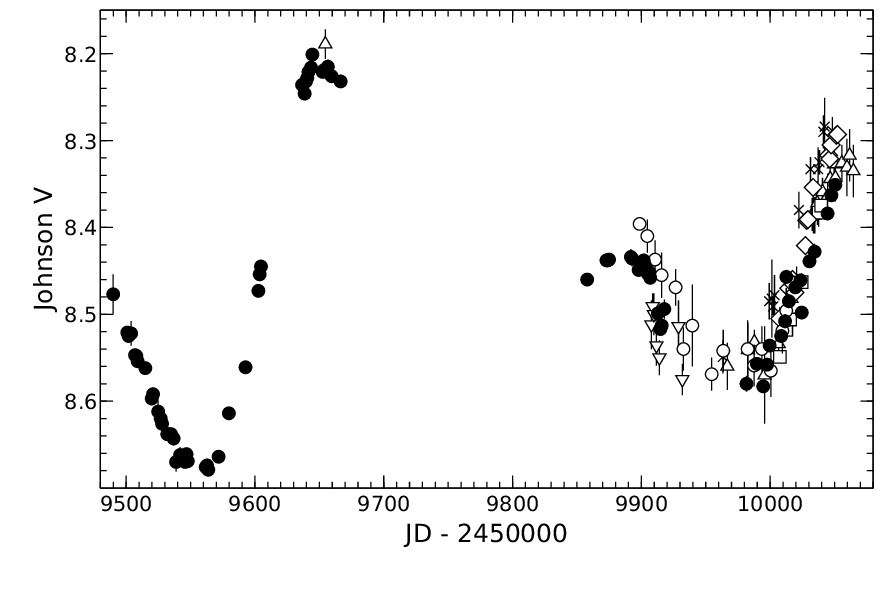}
  \caption{Johnson $V$ photometry of TU Tau.  The symbols refer to different
    observatories as follows: $\bullet$ : DSO; $\circ$: BHU; $\diamond$: CDZ;
    $\square$: LMA; $\vartriangle$: SFOA;
    $\triangledown$: WGR; $\times$: BDG. Some observers provided multiple
    closely spaced observations on a single night.  These were averaged and the
    observational error was taken as the standard deviation of those
    data points.}
  \label{fig:V}
\end{figure}

Figure \ref{fig:Sv} shows a plot of the differential Str\"{o}mgren-$v$
photometry obtained
with the DSO-6 wide-field imager at the Dark Sky Observatory.  The importance
of Str\"{o}mgren-$v$ photometry is that the filter passband is
centered on and
confined to the region of the spectrum where the A-star companion dominates.
The flux in the Johnson-$V$ photometric band, on the other hand, is dominated
by the carbon star (see Figure \ref{fig:fig1}).  It is clear that the
Str\"{o}mgren-{\it v} light curve shows a dramatic and irregular dimming of
the A-type star commencing at about JD2459900.  Figure \ref{fig:V} shows the
Johnson-$V$ light curve during the same time period.

\subsection{Spectroscopy}

Classification resolution spectroscopy ($R \approx 1300$) of TU Tau has been
obtained regularly using the GM spectrograph on the 0.8-m telescope at the
Dark Sky Observatory since October 2021.  The 600 g mm$^{-1}$ grating was
utilized along with an Apogee camera with a blue-enhanced e2v chip
with $27\mu$m pixels (unbinned) to yield a spectral range from
$3800 - 5700$\AA.  These spectra
were reduced with a custom python pipeline which uses functions from astropy
\citep{astropy2013,astropy2018,astropy2022} and
ccdproc \citep{matt_craig_2017_1069648}.  The pipeline employs optimal extraction of the spectra using the
algorithm from \citet{horne86} and detailed modelling of the sky background.
Approximate
flux calibration was carried out through the observation of several
spectrophotometric standards most nights.  The flux calibration was placed on
an absolute scale using the Johnson B photometry from the DSO-6 wide-field
imager.

\begin{table*}
  \caption{Sources of TU Tau Spectroscopy}
  \label{tab:spectra}
  \resizebox{\textwidth}{!}{%
  \begin{tabular}{llllllll}
    \hline
    \hline
    Observatory & Telescope Diam (m) & Spectrograph & Resolution & Spectral Range (\AA) & Location & Observer & Abbreviation\\
    \hline
    Dark Sky Observatory & 0.80 & GM spectrograph & 1300 & 3800 -- 5700 & North Carolina & various & DSO\\
    Vatican Observatory & 1.8 & VATTspec & 3000 & 3750 -- 5500 & Arizona & Christopher Corbally & VATT\\
    LAMOST Observatory & 4.0 & Low Resolution mode & 1800 & 3700 -- 9000 & Hebei, China & various & LAMOST\\
    Lost Gold Observatory & 0.41 & Shelyak ALPY 600 & 500 & 3750 -- 7250 & Arizona & Robert Buchheim & LGO\\
    Observatoire de l'Eridan et & 0.20 & Shelyak ALPY 600 & 530 & 3700 -- 7565 & \'Epaux-B\'ezu, France & Christophe Boussin & OECB\\
    \phantom{O}\phantom{O}de la Chevelure de B\'er\'enice & & & & & & \\
    West Challow Observatory & 0.28 & Shelyak LISA & 1000 & 3900 -- 7400 & Wantage, UK & David Boyd & WCO\\
    Desert Celestial Observatory & 0.51 & Shelyak LISA & 1000 & 3750 -- 7300 & Arizona & Forrest Sims & DCO\\
    Observatorio El Gallinero & 0.31 & Shelyak LISA & 600 & 3900 -- 7500 & Madrid, Spain & David Cejudo Fernandez & OEG\\
    SAROS-1 & 0.25 & Shelyak UVEX & 1400 & 3700 -- 6900 & California  & Jim Grubb & SAROS\\
    Huggins Spectroscopic Observatory & 0.36 & Shelyak LHIRES III & 2000 & 3944 -- 4939 & UK & Jack Martin & HSO\\
   Desert Wing Observatory & 0.36 & Shelyak LHIRES & 14000 & 6482 -- 6632  & Arizona & Albert Stiewing & SALC\\
    Daglen Observatory & 0.36 & Shelyak LHIRES III & 14000 & 6333 -- 6493   & New Mexico & Joe Daglen & DO\\
    DEVO Observatory & 0.36 & Shelyak LISA & 1000 & 3850 -- 7348 & Texas & Keith Shank & DEVO\\
    \hline
  \end{tabular}}
\end{table*}
    
In addition spectra were obtained on the 1.8-m Vatican Observatory Advanced
Technology Telescope (Alice P. Lennon Telescope) employing the VATTspec, a 1
arcsecond slit, a
600 g mm$^{-1}$ grating and a STA0520A back-thinned CCD with 15$\mu$m pixels,
which yield a resolution of $R = 3000$ and a spectral range of
$3750 - 5500$\AA.
The VATTspec spectra were reduced with IRAF \citep{iraf86,iraf93} using
standard procedures, and were approximately flux calibrated via the observation
of a few spectrophotometric standards.

Spectra were
also obtained at a number of amateur observatories listed in Table 3.  These
spectra were reduced using various packages designed for amateur
use, including ISIS
\footnote{\url{http://www.astrosurf.com/buil/isis-software.html}} and
BASS \footnote{\url{https://groups.io/g/BassSpectro}}.  Flux calibration
was carried out
using observations of standard stars and placed on an absolute scale via
Johnson V photometry.

\section{Discussion}

\subsection{Spectral Classification}
\label{sec:class}

During the period covered by this paper (October 2021 -- April 2023), the
spectral type of the carbon-star component of TU Tau (TU Tau A) remained
unchanged even though the Johnson $V$ magnitude varied by over half a magnitude.
We agree with the spectral type reported by \citet{barnbaum96}, C-N4$^+$ C$_2$6.

On the other hand, the spectrum of the A-star component (TU Tau B) underwent
some marked and interesting changes, illustrated in
Figure \ref{fig:classification}.  In addition to the many spectra obtained
at DSO ($R = 1300$) and the observatories listed in Table
\ref{tab:spectra}, some of which have higher resolution than the DSO
spectra, but very restricted spectral ranges, we have four higher resolution
spectra which we can use to derive detailed spectral types for the A-type
component.  Three of these
are from VATTspec ($R = 3000$) and one from LAMOST ($R = 1800$).  The VATTspec
spectra were obtained on 10/25/2021, 01/21/2022 and 01/12/2023.  The LAMOST
spectrum was acquired on 12/20/2013.  We will begin with detailed comments on
the VATTspec spectrum obtained on 10/25/2021, as it shows the least pronounced
peculiarities.

Our spectral types are based on the standard classification criteria for A-type
stars \citep{gray09}, which include the widths and strengths of the Balmer
lines, the strength of the Ca II K-line, and the strength of the general
metallic-line spectrum.

These classifications were carried out in comparison with the standard
stars in the MKCLASS library {\tt libr18}\footnote{\url{https://www.appstate.edu/~grayro/mkclass/}}\citep{gray14}.
All classifications reported
here were carried out manually and did not utilize the automatic spectral
classification program MKCLASS.  However, we did
employ the manual classification program {\tt xclass} which utilizes
{\tt libr18} and can interpolate between the standards in that library.

\vspace{0.2cm}

\noindent VATT 10/25/2021: The hydrogen-line wings as well as the general
metallic-line
spectrum are consistent with a spectral type of A2 III-IV for this spectrum
of TU Tau B.  However, there are a few ways in which this spectrum deviates
from the standards.   The Sr II $\lambda 4077$ and $\lambda 4216$ lines are
both somewhat
strong compared to the A2 III-IV standard.  Likewise, the Fe II $\lambda 4233$
line
is stronger in this spectrum than in the A2 III-IV standard.  A strong
Fe II $\lambda 4233$ line can indicate the presence of circumstellar gas
and is a
criterion for classifying a star as a ``shell'' star
\citep[see][and below]{gray09}.  In this case the Fe II $\lambda 4233$ line
is not strong
enough to warrant a full ``shell'' classification. Another peculiarity is
that the Ca II K-line is stronger than for a normal A2 star; it is
intermediate in strength between A2 and A3.

\begin{figure}
  \includegraphics[width=\columnwidth]{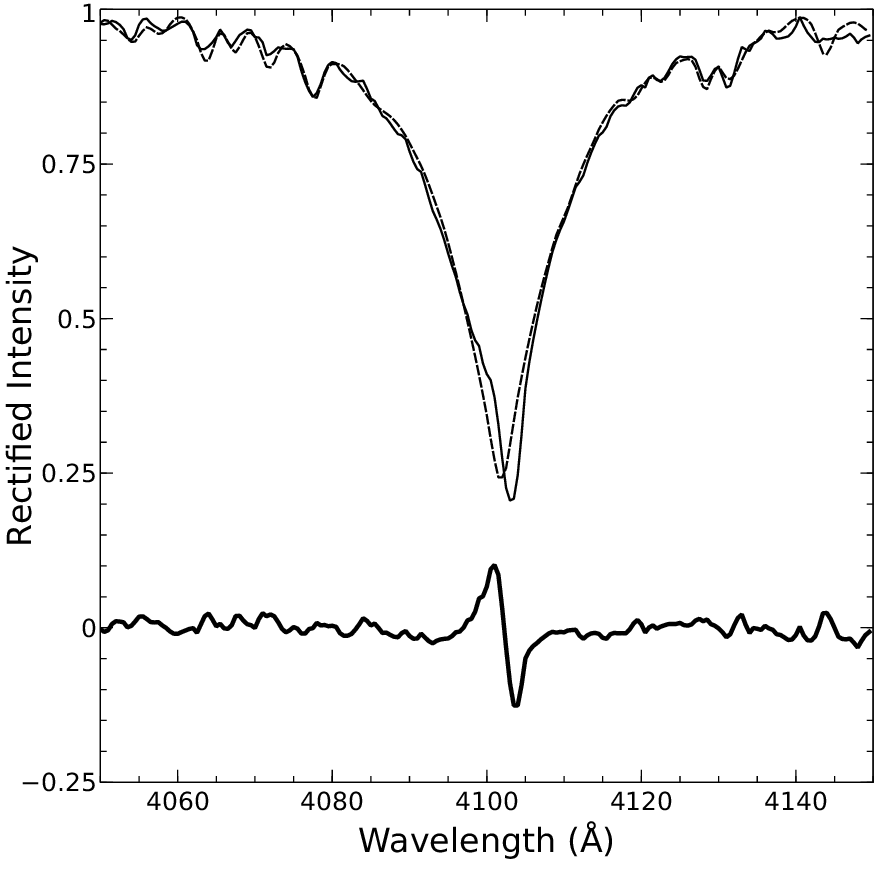}
  \caption{The solid line shows the H$\delta$ profile of the 10/25/2021
    VATT spectrum of TU Tau B.  The dashed line shows the H$\delta$ profile
    of the A2 III-IV standard.  Both stars are represented in their rest
    frames.  The red-shifted core of TU Tau B is easily discerned.  The bold
    solid line shows the difference spectrum (TU Tau B
    $-$ A2 III-IV standard).  What is revealed is an inverse P-Cygni profile,
    a spectroscopic signature of accretion.  See Sections \ref{sec:class} and
    \ref{sec:analysis} for further details.}
  \label{fig:pcyg}
\end{figure}

One of the most notable ways this spectrum deviates from the A2 III-IV
standard, however, is in the hydrogen-line profiles.  The wings of the
hydrogen lines are in
excellent agreement with the A2 III-IV standard, but all of the visible
Balmer lines show narrow, slightly deep, and redshifted cores (see
Figure \ref{fig:pcyg}).  A difference spectrum (TU Tau B $-$
A2 III-IV standard) for the H$\delta$ line shows that this redshifted core
arises from an inverse P-Cygni profile (see Figure
\ref{fig:pcyg}).  An inverse P-Cygni profile is an indication of active
accretion of gas, and so is consistent with the shell nature of the star
we deduced above.  We will discuss the implications of this accretion
in more detail in Section \ref{sec:analysis}.  The Ca II K-line, in addition,
shows
a marked asymmetry and the core is also shifted to the red, probably also
a consequence of accretion.  We
assign a spectral type A2 III-IVp kA2.5 (Sr) ((shell)) for this spectrum.

\begin{figure}
  \includegraphics[width=\columnwidth]{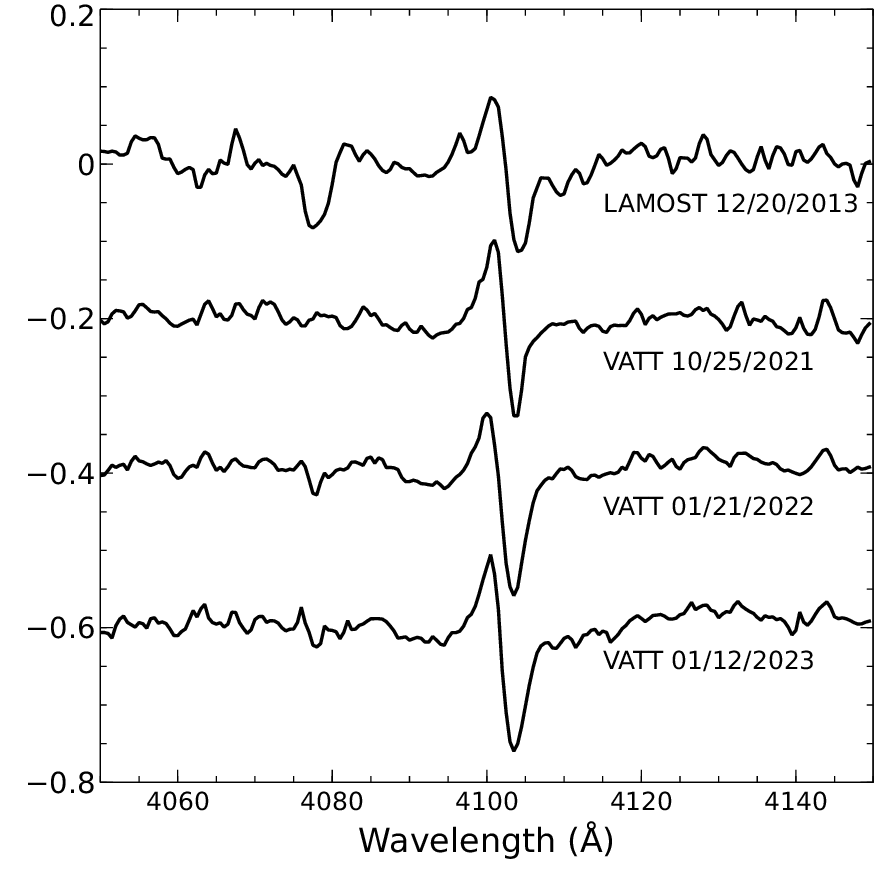}
  \caption{Difference spectra for the LAMOST spectrum and the three VATT
    spectra at H$\delta$ (see Figure \ref{fig:pcyg}).  Note that the inverse
    P Cygni profiles are all similar although some variation is visible.
    In the LAMOST difference spectrum,
    the feature at $\lambda 4077$ arises from the much stronger Sr II
    $\lambda 4077$ line in the LAMOST spectrum compared to the standard
    A2 III-IV spectrum.}
  \label{fig:pcyg2}
  \end{figure}
    
\vspace{0.2cm}

\noindent VATT 01/21/2022: The hydrogen-line wings and the general
metallic-line spectrum are in good agreement with a spectral type of A2 III-IV.
The metallic-line spectrum, however, shows more peculiarities than the
10/25/2021 spectrum.  A few lines/blends appear weak (4143\AA\ and 4187\AA\ in
particular, both dominated by Fe I).  However, both Sr II $\lambda 4077$ and
$\lambda 4216$ are
stronger than normal.  The Fe II $\lambda 4233$ line is exceptionally
strong.  It is only in
A supergiants and shell stars that Fe II $\lambda 4233$ reaches this strength
\citep[see][]{gray09}.  This suggests a ``shell'' classification for this
spectrum.  Such a classification is normally confirmed by examining the
strength of the Fe II and Ti II lines in the ``Fe II, Ti II forest''
between H$\gamma$ and 4500\AA, and the strengths of the Fe II multiplet 42
lines -- $\lambda\lambda$4924, 5018, 5169 \citep[see][]{gray09}, but those
spectral regions are obscured by the carbon star.  However, two lines in the
violet, Ti II $\lambda 3900$ and Fe II, Ti II $\lambda 3913$, also enhanced in
A-type shell stars, are unusually strong in this spectrum.
This confirms our ``shell'' classification for this spectrum.  The
hydrogen-line profiles show the same asymmetries seen in the 10/25/2021
spectrum and a difference spectrum shows the same inverse P-Cygni profile
(see Figure \ref{fig:pcyg2}).
The Ca II K line likewise is asymmetrical and redshifted.  In
this spectrum the total strength of the Ca II K-line is intermediate to the
A3 and A4 standards.
This yields the spectral type  A2 III-IVp kA3.5 Sr shell.

\vspace{0.2cm}

\noindent VATT 01/12/2023:  This spectrum is similar to the 01/21/2022
spectrum, showing the same peculiarities and asymmetries.  However, the
shell lines noted above in the 01/21/2022 spectrum are somewhat weaker.
This yields the spectral type A2 III-IVp kA3.5 Sr (shell).

\vspace{0.2cm}

\noindent LAMOST 12/20/2013:  The LAMOST spectrum is the most peculiar of
the four.  The hydrogen-line wings agree with the A2 III-IV standard. However,
the general metallic-line spectrum is stronger, and is more similar in
strength to an A3 star than an A2.  A number of lines are very strong.  The
Sr II $\lambda 4077$ line is very strong, as is Sr II $\lambda 4216$ (see
Table \ref{tab:eqw}).
Lines of other {\it s}-process elements, Y II $\lambda 4374$ and Ba II
$\lambda 4554$ (although
that line may have a significant contribution from the carbon star) are
also very strong.  Interestingly, Ca I $\lambda 4226$, a resonance line
of neutral calcium
is also very strong.  The Fe II $\lambda 4233$ shell line has a broad
profile but its core
is not as deep as in the VATT 01/21/2022 spectrum, so the equivalent width
recorded in Table \ref{tab:eqw} is uncertain.  The hydrogen lines also show
the same asymmetries displayed by the three VATT spectra, and a difference
spectrum shows a clear inverse P-Cygni profile (Figure \ref{fig:pcyg2}).
The Ca II K-line has a
strength similar to that of the A3 standard, but in this spectrum it has a
symmetrical profile and is not redshifted.  These considerations yield the
spectral type:
A2 kA3mA3 III-IVp SrYBa:

\begin{table*}
  \caption{Equivalent Widths (m\AA) of Selected lines in TU Tau Spectra}
  \label{tab:eqw}
  \begin{tabular}{llllll}
    \hline
    \hline
    Spectrum & Fe II, Ti II $\lambda 3913$ & Sr II $\lambda 4077$ &
    Sr II $\lambda 4216$ & Ca I $\lambda 4226$ & Fe II $\lambda 4233$ \\
    \hline
    LAMOST (107807030) & blended & 594    & 582      & 837       & 422: \\
    VATT 10/25/21 & 206   & 215        & 217        & 245       & 302 \\
    VATT 01/21/22 & 302   & 233        & 211        & 213       & 444 \\
    VATT 01/12/23 & 264   & 197        & 249        & 225       & 343 \\
    \hline
    A2 III-IV std & 164  & 184        & 139        & 211       & 226 \\
    \hline
  \end{tabular}
\end{table*}

\vspace{0.2cm}

The most dramatic change in the spectrum of TU Tau, however, was the
disappearance of the A-star beginning in February 2023.  The dimming of this
component was gradual and irregular and characterized by one or more brief
reappearances.  This phenomenon is discussed in Section \ref{sec:eclipse}.

\subsection{Basic Physical Parameters}
\label{sec:param}

Our spectral types and Str\"{o}mgren-$v$ photometry of the A-type companion
allow a rough calculation of the basic physical parameters for the
companion star, including the luminosity, mass, radius, and the
effective temperature.  Recall that the carbon star contributes negligible
flux in the S-{\it v} band.  Our intermediate goal is to use the
S-{\it v} photometry
for TU Tau B to derive an estimate for the absolute magnitude, $M_V$, of
TU Tau B in the Johnson-$V$ band.  This requires a number of steps.  First,
an examination of Figure \ref{fig:Sv} indicates that the brightness of the
A-star companion varies.  We will argue in Section \ref{sec:eclipse} that
the A-star companion is being obscured by clumps and shells of dust in the
outflow from the carbon star.  From that figure we can see that the A-star
is brightest on the night of JD2459857 and we assume that on that night the
S-{\it v} magnitude represents the minimally obscured brightness of the A-star,
modified only by the interstellar reddening in the direction of TU Tau
plus a possible internal reddening along our line of sight to TU Tau B in the
TU Tau system itself. To begin with we will ignore the internal reddening,
but will attempt to estimate it later.

We begin by placing the S-{\it v} photometry for TU Tau B on the night of
JD2459857 on the standard system.  This is done via the star
HD~38334, an A3 star about $15'$ from TU Tau which was observed on the same
exposures of TU Tau taken with the wide-field DSO-6 imager.  HD~38334 has
been observed on the Str\"{o}mgren system, with the following magnitudes:
$V = 7.96$, $b-y = 0.118$, and $m_1 = 0.180$ \citep{perry82}.  Using the
definition of $m_1 = (v-b) - (b-y)$  (where $v$ corresponds to what we
  are calling S-$v$ in this paper) and the fact that
Str\"{o}mgren $y$ is transformed to be on the same system as Johnson $V$, we
determine that S-$v$(HD~38334)$ = 8.38 \pm 0.01$.  On the night of
  JD2459857, the differential magnitude in the S-{\it v} band between TU Tau B
  and HD 38334 was $3.340 \pm 0.028$ mag, yielding S-$v$(TU Tau B)$ = 11.72 \pm 0.03$.

Note: in the calculations that follow, we have used a Monte Carlo method for the
propagation of errors \citep{Crowder2020}, and have assumed that the errors
of the observed quantities have gaussian distributions.

The interstellar reddening at the distance and position of TU Tau can be
estimated from the 3-D reddening map of \citet{green19}: $E(g-r) = 0.28^{+0.03}_
{-0.04}$.  This can be converted to the color excess in the Johnson system
using the formula in that reference: $E(B-V) = 0.981E(g-r)$.  This yields
$E(B-V) = 0.27^{+0.03}_{-0.04}$.  The corresponding absorption in
  the S-$v$ band
  can be estimated by interpolating in the tablulation of $A(\lambda)/A(V)$
  in Table 21.6
  of \citet{cox00}.  $A(V) = R_V E(B-V)$ where we take $R_V$, the ratio of total
  to selective absorption $= 3.1$, the typical value for the diffuse
  interstellar medium \citep[cf.][]{cardelli89}. $A(V)$ is the total absorption
  in magnitudes in the $V$-band, whereas $A(\lambda)$ is the corresponding
  total absorption in a band centered at wavelength $\lambda$.  For the S-$v$
  band, $\lambda = 0.41\mu {\rm m} = 4100$\AA.  The reddening-free magnitude
  in the S-$v$ band, S-$v_0$, is given by S-$v - A(0.41\mu{\rm{m}})$.  This
yields S-$v_0$(TU Tau B)$ = 10.54 \pm 0.18$ as the
reddening-free S-$v$ magnitude of TU Tau B (note again that at this point we are
assuming that there is no internal reddening in the TU Tau system.  This is
unlikely, and we consider that question below).  To use this to deduce the
reddening-free Johnson-$V_0$ magnitude for TU Tau B, we must derive
typical reddening-free
Str\"{o}mgren colors for an A2 III-IV star.  This can be accomplished by
pulling out lightly reddened stars from \citet{gray87} with similar spectral
types to TU Tau B (A2 IV, A2 III-IV, A2 III).  We find, for a typical
A2 III-IV star, $(b-y)_0 = 0.041 \pm 0.016$ and $m_0 = 0.153 \pm 0.013$.  This
gives $V_0$(TU Tau B)$ = 10.31 \pm 0.18$.  The {\it Gaia} DR3 parallax
\citep[$0.8482 \pm 0.0341$ mas;][]{gaia2016b,gaia2022k}
for TU Tau yields $M_V$(TU Tau B)$ = -0.05 \pm 0.21$.

\begin{figure}
  \includegraphics[width=\columnwidth]{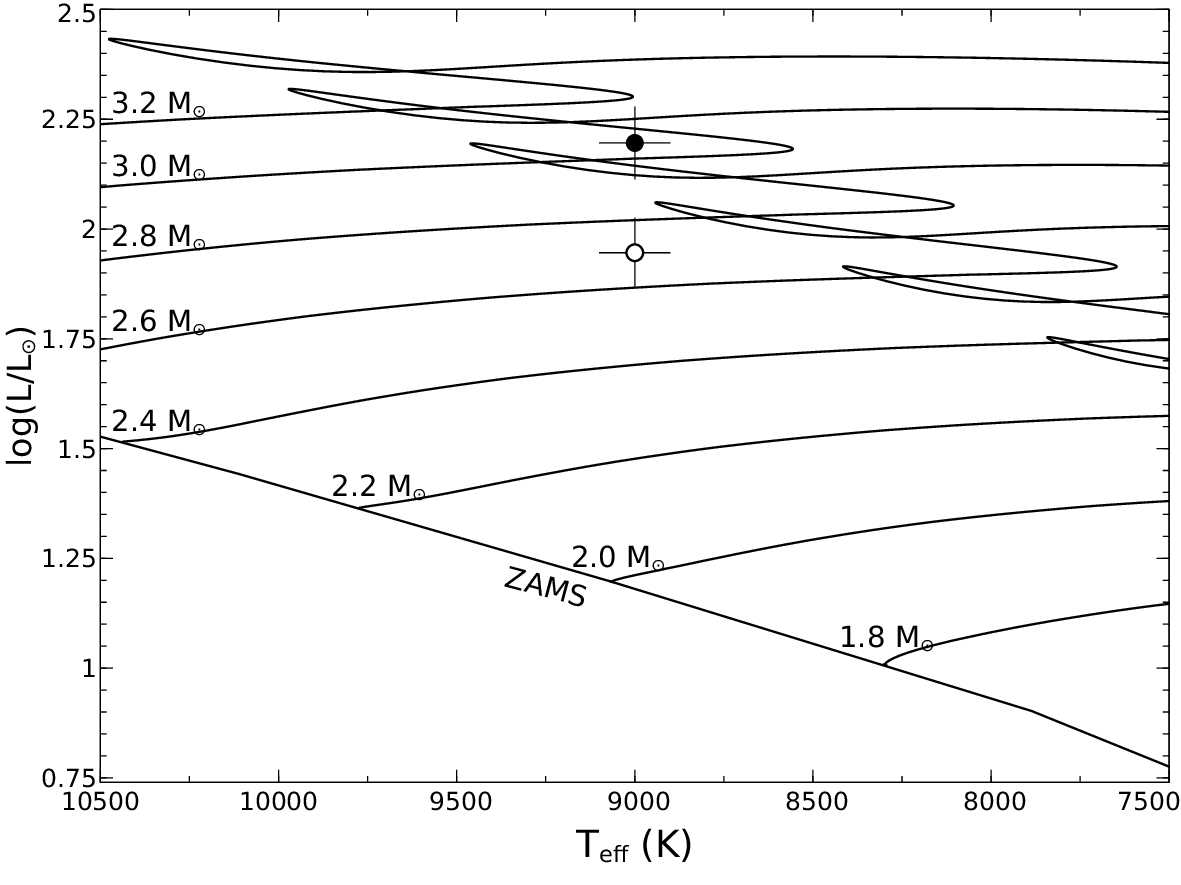}
  \caption{A theoretical HR diagram based on the $Z = 0.02$ evolutionary
    tracks of \citet{bressan12} shows the position of the two solutions for
    TU Tau B discussed in Section \ref{sec:param}.  The point indicated by
    the open circle assumes no internal reddening in the TU Tau system.  The
    filled circle indicates the solution for an internal reddening of
    $E(B-V)_{\rm internal} = 0.14$ (see text).}
  \label{fig:HR}
\end{figure}

The spectral-type, effective temperature calibration of \citet{cox00} agrees
with that of \citet{gray09} in assigning $T_{\rm eff} = 9000$K for an A2 III-IV
star.  We estimate, from the precision of our spectral type, that the error
in that determination is $\pm 100$K.  The bolometric correction from
\citet{flower96} yields $\log(L/L_\odot) = 1.946 \pm 0.080$ for the A-star
companion.  The solar-abundance ($Z = 0.02$)
evolutionary tracks of \citet{bressan12} imply
$M$(TU Tau B)$ = 2.7 \pm 0.1 M_\odot$ (see Figure \ref{fig:HR}).
This finally yields $\log(g) = 3.70 \pm 0.08$ and $R = 3.91 \pm 0.37 R_\odot$.

Spectral synthesis based on {\tt ATLAS9} models \citep{castelli04} and the LTE
spectral synthesis program SPECTRUM\footnote{\url{https://www.appstate.edu/~grayro/spectrum/spectrum.html}} \citep{gray94}, yields a best fit for the
wings of the Balmer
lines and lines of ionized species (Ti II, Fe II) at $\log(g) = 3.5$.  This
implies a slightly higher luminosity than calculated above which indicates
that the internal reddening in the TU Tau system is not negligible.
Setting $E(B-V) = 0.27 + E(B-V)_{\rm internal}$ and iterating to convergence
gives $E(B-V)_{\rm internal} = 0.14$, $\log(L/L_\odot) = 2.196 \pm 0.083$,
$\log(g) = 3.50 \pm 0.09$, $M = 3.05 \pm 0.2 M_\odot$ and $R = 5.18 \pm 0.51
R_\odot$.

 It is important to note that in principle this calculation could begin
  with the S-$v$ magnitude for TU Tau B {\it on any night}.  The choice
  of JD2459857 simply {\it minimizes} the internal reddening that we calculate
  for the line of sight to TU Tau B in the TU Tau system.

The first estimate of the temperature and luminosity of TU Tau B (assuming no
internal reddening) places it in the upper main-sequence band (see Figure
\ref{fig:HR}).  The second
estimate (with $E(B-V)_{\rm internal} = 0.14$) places the star near
to the point of hydrogen exhaustion.

\subsection{The ``Eclipse'' of TU Tau}
\label{sec:eclipse}

\begin{figure}
  \includegraphics[width=\columnwidth]{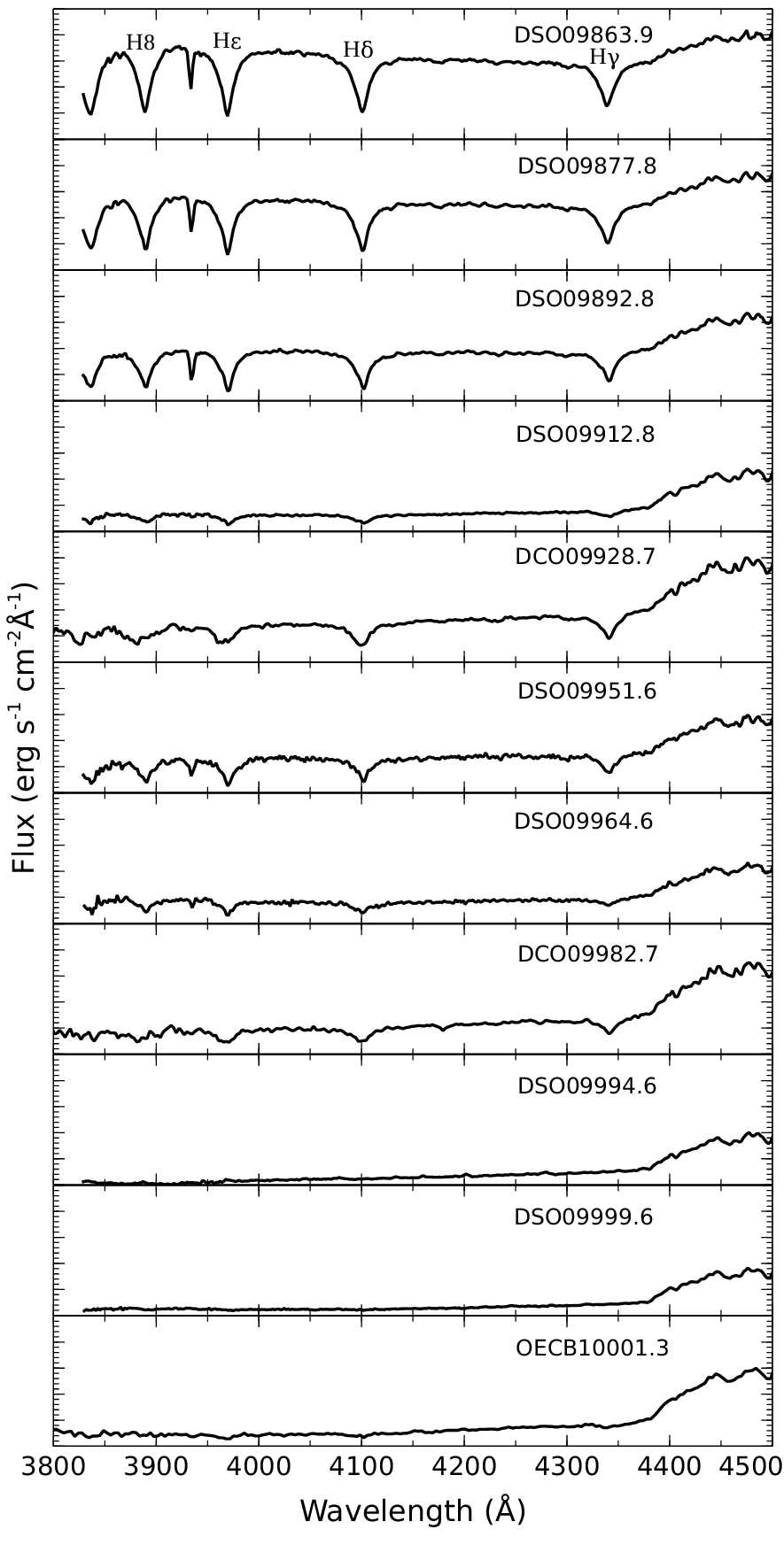}
  \caption{A montage of spectra of TU Tau in the blue-violet showing the
    region of the spectrum where the A-type companion is visible.  In the
    first panel the Balmer lines (labeled) originating in that companion
    are clearly visible.  The panel label indicates the origin of the
    spectrum (DSO: Dark Sky Observatory;
    DCO: Desert Celestial Observatory -- Forrest Sims; OECB: Observatoire de
    l'Eridan et de la Chevelure de B\'er\'enice -- Christophe Boussin)
    and the Julian date, in the format 2460000 - JD.  Each panel is scaled
    identically in
    flux, from 0 to $2.5 \times 10^{-13}$ erg s$^{-1}$ cm$^{-2}$\AA$^{-1}$.  The montage
    begins with a spectrum
    taken on Oct 11, 2022, showing the A-type companion outside of the
    ``eclipse''.  Further spectra track the irregular dimming of the
    A-type companion through the last spectrum in the montage, obtained on
    Feb 25, 2023.
    Note the complete disappearance of the A-type companion between
    JD2459994 and 2459999.  In the last spectrum close inspection reveals
    very weak absorption features at H$\delta$ and H$\epsilon$.
    The montage continues in Figure \ref{fig:spectra2}.}
  \label{fig:spectra1}
\end{figure}

\begin{figure}
  \includegraphics[width=\columnwidth]{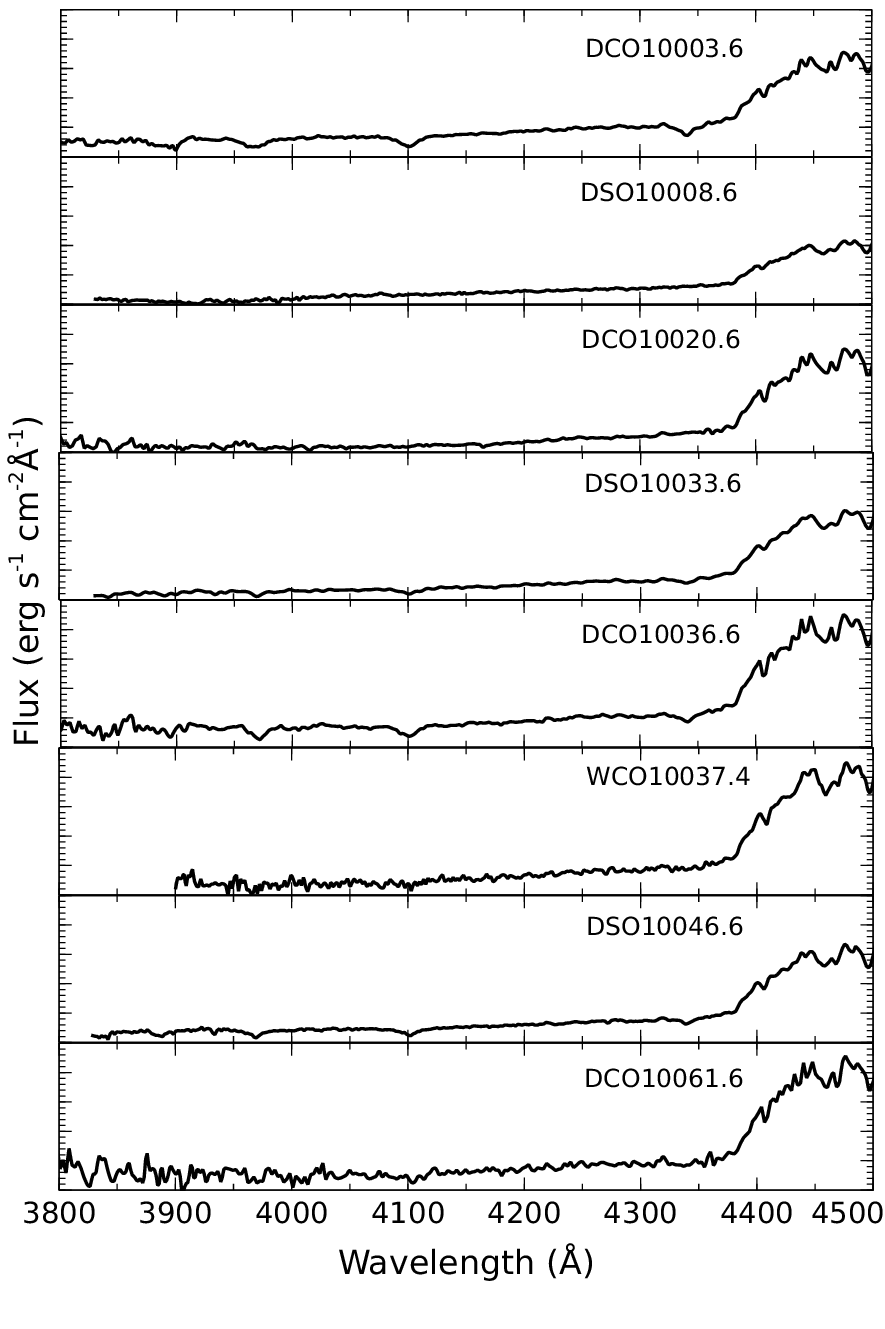}
  \caption{A continuation of the montage in Figure \ref{fig:spectra1}.  The
    plots are scaled identically to that figure.  The observatory codes are
    the same as that figure, except with the addition of WCO: West Challow
    Observatory -- David Boyd.  The A-type star remains faint in these
    spectra, although the visibility of that companion, as judged by the
    strength of the Balmer lines, is variable. Note the partial recovery in
    the visibility of the A-type companion in panel DCO10036.6.  However, a
    fraction of a day later (panel WCO10037.4), the companion is not detected.
    By the end of the observing season (the end of April, 2023), the ``eclipse''
    event is clearly still in progress (panel DCO10061.6).}
  \label{fig:spectra2}
\end{figure}

The S-{\it v} light curve (Figure \ref{fig:Sv}) and the montage of blue-violet
spectra (Figures \ref{fig:spectra1} and \ref{fig:spectra2}) document an
unusual dimming, disappearance and multiple partial reappearances of the
A-star companion which started about JD2459905.  It is clear from those
figures that this dimming event was not the result of a normal eclipse, as
the A-star disappeared and partially reappeared multiple times.  By the end
of the observational season, this event was still ongoing, with the A-type
companion still quite faint.

The most plausible explanation for this unusual event is obscuration of the
A-type companion by nonhomogeneities in the outer atmosphere of the carbon
star, most probably related to dust formation in outflows.  That the A-type
companion is being obscured by dust is tentatively supported by the
observation (see Figures \ref{fig:spectra1} and \ref{fig:spectra2}) that
the energy distribution of that star appears to get redder as it gets fainter.
Two scenarios seem equally plausible, as we
know very little about the orbital dynamics of the binary system.  The first
scenario posits that the orbit of the A-type companion is inclined by an
angle such that a grazing eclipse occurs during which the observer sees the
A-type companion through the extended atmosphere of the carbon star. Dusty
shells and other nonhomogeneities in the extended atmosphere are the
cause of the ``flickering'' of the A-type companion that we have observed.
Under this scenario, we should expect these dimming events to be periodic
with the (currently unknown) orbital period of the A-type companion.

The alternative scenario is that the orbital period of the A-type star is
sufficiently long so that we are not seeing the companion moving through
or behind nonhomogeneities in the extended atmosphere of the carbon star.
Rather, those nonhomogeneities are moving past the A-type star.  
Thus, under this scenario, the flickering of the A-type companion is caused by
nonhomogeneities in the outflow from the carbon star passing in front of
the A-type companion from the viewpoint of the observer.  This suggests
that these dimming events should occur at random intervals.

At the present time we cannot decide between these two scenarios, as there are
only very sparse observational data for TU Tau B prior to this study.  One
data point we do have is provided by the spectrum of TU Tau published by
\citet{barnbaum96} and acquired on Feb 15, 1995 (kindly provided in a
private communication from Dr. Barnbaum).  The A-type companion is not
visible in that spectrum.  \citet{richer72} obtained a blue-violet spectrum
of TU Tau that does show the A-type star at some point before the publication
of that paper.  Apart from that, we have been unable to find any other data
that pertains to the visibility of the A-star companion in the literature.
Archival spectra of TU Tau would be of value in deciding between these two
scenarios.  On the other hand, the S-{\it v} light curve of TU Tau (Figure
\ref{fig:Sv}) does show the A-type companion undergoing a slow brightening
between JD2459489 and JD2459666.  During that time the Johnson-$V$ light curve
(Figure \ref{fig:V}) shows the carbon star going through a deep minimum.
This suggests that during this period the A-type companion was ``recovering''
from a similar dimming event, perhaps occuring around JD2459300, hinting at
a ``period'' of about 700 days.  Further observations are required.

With our estimate for the $V$ magnitude of TU Tau B derived in Section
\ref{sec:param}, it is possible to
calculate the magnitude change for the entire system during a complete
disappearance of the A-star component.  This works out to be
$\Delta V \approx 0.1$.
This is consistent with the magnitude drop seen in Figure \ref{fig:V} at the
beginning of the A-star disappearance (about JD2459900).   

\section{Analysis and Conclusions}
\label{sec:analysis}

The results presented in this paper demonstrate that the A-type component
of the TU Tau binary is interacting in a number of ways with the complex
environment created by outflows from the carbon star.  Not only is the
ultraviolet flux from the A-star inducing complex photochemistry in
the carbon star outflow (see discussion in the Introduction), it is now
clear, from the inverse P-Cygni profiles
seen in the difference spectra illustrated in Figures \ref{fig:pcyg} and
\ref{fig:pcyg2}, that the A star is actively accreting gas from the outflow.
That accretion has interesting implications for the evolution of the A-star.
C-N carbon stars, like the primary in the TU Tau system, have overabundances
of carbon and {\it s}-process elements in their envelopes and atmospheres,
dredged up by deep
convective currents from a layer that has undergone helium fusion.  Those
enhanced abundances will also be present in the outflows from the carbon
star.  It is clear, especially from the LAMOST spectrum of the A-type
companion which shows very strong lines of {\it s}-process elements, that
the material accreting onto the surface of the A star is also enriched in
those elements.  This is exactly the scenario proposed by \citet{mcclure80}
to explain the creation of Barium giants.  Barium giants are G- and K-type
giants that show large overabundances of {\it s}-process elements at a
stage of evolution well before those elements are produced during shell helium
burning.  It is believed that those overabundances result from mass transfer
of {\it s}-process enriched material from the carbon star to (what will
become) the G/K-type companion.. As far as can be determined, all Barium
giants have white dwarf companions \citep{bohm-vitense00}, thought to be
remnants of the former carbon star. As \citet{bohm-vitense00} argued (see
Section \ref{sec:intro}), those Barium giants must have been contaminated 
while they were still on the main sequence (e.g. before hydrogen exhaustion).
Since  most Barium giants have masses $M \ge 2.0M_\odot$ \citep{escorza17},
those stars were B- or
A-type stars at the time of contamination.  We have shown that the
mass of the A-star component of TU Tau is $\approx 3M_\odot$.  We have also
shown clear
evidence of active accretion of {\it s}-process enriched material onto the
A-star companion.  It thus appears that TU Tau B is a prime candidate
for a Barium
giant in-the-making.  

There remain some questions about the nature of this accretion.  Is the
A-star accreting material directly from the outflow from the carbon star, or
has the carbon star filled its Roche lobe?  This can be answered only when we
learn more about the orbital dynamics of the system.  We unfortunately are
not able to observe the A-star at H$\alpha$ or H$\beta$, as the profiles of
those lines might be diagnostic.  However, observation of the A-star spectrum
at higher resolution may help to elucidate the details of the accretion
process.  Analysis of the inverse P-Cygni profiles indicates that the
velocity of the accretion inflow is on the order of 200km/s.  The shell lines,
in particular Fe II 4233, are presumably formed via photoionization of the
accreting material by ultraviolet photons from the A-type star itself.
Analysis of the profile of that line in higher resolution spectra
may reveal details of the
velocity structure of the inflow and temporal changes in the accretion.
Recall the
rather broad, ill-defined profile of that line in the LAMOST spectrum (see
Section \ref{sec:class}) in contrast to the more narrow profiles in the
VATT spectra.

Another open question arises from the relative strengths of the Sr II
$\lambda 4077$ and $\lambda 4216$ lines.  These are resonance lines
of the strontium ion, and belong to the same multiplet.  Because the
$gf$ (statistical weight $\times$ oscillator strength) value of $\lambda 4077$
exceeds that of $\lambda 4216$ by a factor of two, the ratio of the
equivalent widths of those lines,
EqW(Sr II $\lambda 4077$)/EqW(Sr II $\lambda 4216$) should be $> 1$.  In
the case of the A2 III-IV standard, that ratio $\approx 1.32$
(see Table \ref{tab:eqw}).  However, that same ratio ranges from 0.79 to
1.10 in TU Tau.  What that means
is that if it is assumed those lines are formed exclusively in the
photosphere of the star, those lines will give inconsistent strontium
abundances.  This will be the case even if strontium is distributed
nonuniformly across the surface of the star.  However, if the strontium lines
are formed at least in part in the accreting material, then the peculiar
ratios might be explained either through the velocity structure of that
accreting material or via a strontium abundance gradient in that material.
This is because those two lines will be formed at different heights in that
accreting flow, with the $\lambda$4077 line formed higher up than the
$\lambda$4216 line.  Only detailed modeling based on higher resolution
and higher signal-to-noise spectra will be able to resolve this
question.

We have suggested that the TU Tau system represents a stage in the evolutionary
process that will result in a Barium giant star.  Even if that is not the
case, the TU Tau system is of great astrophysical interest, as the A-type
companion acts as a probe of the outflows and outer atmosphere of the carbon
star as shown graphically by the remarkable ``flickering'' of the A-type
companion
documented in Figure \ref{fig:Sv} and the montage in Figures
\ref{fig:spectra1} and \ref{fig:spectra2}.  For that reason alone, TU Tau
needs to be studied at higher resolution and over a longer period of time.

\section*{Acknowledgements}
We are grateful for the comments of an anonymous reviewer which have
significantly improved this paper.
This work has made use of data from the European Space Agency (ESA) mission
{\it Gaia} (\url{https://www.cosmos.esa.int/gaia}), processed by the {\it Gaia}
Data Processing and Analysis Consortium (DPAC,
\url{https://www.cosmos.esa.int/web/gaia/dpac/consortium}). Funding for the DPAC
has been provided by national institutions, in particular the institutions
participating in the {\it Gaia} Multilateral Agreement.

This work has used observations acquired at the Vatican Observatory Advanced Technology Telescope, Mt. Graham, Arizona.

This paper has made use of a spectrum from the Guoshoujing (LAMOST) telescope.  The Guoshoujing Telescope (the Large Sky Area Multi-Object Fiber Spectroscopic Telescope LAMOST) is a National Major Scientific Project built by the Chinese Academy of Sciences. Funding for the project has been provided by the National Development and Reform Commission. LAMOST is operated and managed by the National Astronomical Observatories, Chinese Academy of Sciences. 

This work made use of Astropy (\url{http://www.astropy.org}): a community-developed core Python package and an ecosystem of tools and resources for astronomy \citep{astropy2013, astropy2018, astropy2022}. 

This research made use of ccdproc, an Astropy package for
image reduction \citep{matt_craig_2017_1069648}.

This research made use of Photutils, an Astropy package for
detection and photometry of astronomical sources
\citep{larry_bradley_2023_7946442}.

This research was made possible through the use of the AAVSO Photometric All-Sky Survey (APASS), funded by the Robert Martin Ayers Sciences Fund and NSF AST-1412587.

\software{Astropy \citep{astropy2013, astropy2018, astropy2022}}

\software{ccdproc \citep{matt_craig_2017_1069648}}

\software{SPECTRUM \citep{gray94}}

\bibliography{grayro}{}

\begin{thebibliography}{}
\expandafter\ifx\csname natexlab\endcsname\relax\def\natexlab#1{#1}\fi

\bibitem[{Barnbaum {et~al.}(1996)Barnbaum, Stone, \& Keenan}]{barnbaum96}
Barnbaum, C., Stone, R. P.~S., \& Keenan, P.~C. 1996, ApJS, 105, 419

\bibitem[{Boersma {et~al.}(2006)Boersma, Hony, \& Tielens}]{boersma06}
Boersma, C., Hony, S., \& Tielens, A. G. G.~M. 2006, A\&A, 447, 213

\bibitem[{Bohm-Vitense {et~al.}(2000)Bohm-Vitense, Carpenter, Robinson, Ake, \&
  Brown}]{bohm-vitense00}
Bohm-Vitense, E., Carpenter, K., Robinson, R., Ake, T., \& Brown, J. 2000, ApJ,
  533, 969

\bibitem[{Bradley(2023)}]{larry_bradley_2023_7946442}
Bradley, L. 2023, astropy/photutils: 1.8.0, v.1.8.0,  Zenodo,
  doi:10.5281/zenodo.7946442

\bibitem[{Bressan {et~al.}(2012)Bressan, Marigo, Girardi, Salasnich, Dal~Cero,
  Rubele, \& Nanni}]{bressan12}
Bressan, A., Marigo, P., Girardi, L., {et~al.} 2012, MNRAS, 427, 127

\bibitem[{Cardelli {et~al.}(1989)Cardelli, Clayton, \& Mathis}]{cardelli89}
Cardelli, J.~A., Clayton, G.~C., \& Mathis, J.~S. 1989, ApJ, 345, 245

\bibitem[{Castelli \& Kurucz(2004)}]{castelli04}
Castelli, F., \& Kurucz, R.~L. 2004, New Grids of ATLAS9 Model Atmospheres, , ,
  arXiv:astro-ph/0405087

\bibitem[{Cox(2000)}]{cox00}
Cox, A.~N. 2000, Allen's Astrophysical Quantities, 4th edn. (New York, New
  York: Springer-Verlag)

\bibitem[{Craig {et~al.}(2017)Craig, Crawford, Seifert, Robitaille, Sip{\H
  o}cz, Walawender, Vin{\'{\i}}cius, Ninan, Droettboom, Youn, Tollerud, Bray,
  Walker, Janga, Stotts, G{\"u}nther, Rol, Bach, Bradley, Deil, Price-Whelan,
  Barbary, Horton, Schoenell, Heidt, Gasdia, Nelson, \&
  Streicher}]{matt_craig_2017_1069648}
Craig, M., Crawford, S., Seifert, M., {et~al.} 2017, astropy/ccdproc:
  v1.3.0.post1, , , doi:10.5281/zenodo.1069648

\bibitem[{Crowder {et~al.}(2020)Crowder, Delker, Forrest, \&
  Martin}]{Crowder2020}
Crowder, S., Delker, C., Forrest, E., \& Martin, N. 2020, Monte Carlo Methods
  for the Propagation of Uncertainties (Cham: Springer International
  Publishing), 153--180

\bibitem[{Escorza {et~al.}(2017)Escorza, Boffin, Jorissen, Van~Eck, Siess,
  Van~Winckel, Karinkuzhi, Shetye, \& Pourbaix}]{escorza17}
Escorza, A., Boffin, H. M.~J., Jorissen, A., {et~al.} 2017, A\&A, 608, 100

\bibitem[{Flower(1996)}]{flower96}
Flower, P.~J. 1996, ApJ, 469, 355

\bibitem[{{Gaia Collaboration} {et~al.}(2016){Gaia Collaboration}, {Prusti,
  T.}, {de Bruijne, J. H. J.}, {Brown, A. G. A.}, {Vallenari, A.}, {Babusiaux,
  C.}, {Bailer-Jones, C. A. L.}, {Bastian, U.}, {Biermann, M.}, {Evans, D. W.},
  {Eyer, L.}, {Jansen, F.}, {Jordi, C.}, {Klioner, S. A.}, {Lammers, U.},
  {Lindegren, L.}, {Luri, X.}, {Mignard, F.}, {Milligan, D. J.}, {Panem, C.},
  {Poinsignon, V.}, {Pourbaix, D.}, {Randich, S.}, {Sarri, G.}, {Sartoretti,
  P.}, {Siddiqui, H. I.}, {Soubiran, C.}, {Valette, V.}, {van Leeuwen, F.},
  {Walton, N. A.}, {Aerts, C.}, {Arenou, F.}, {Cropper, M.}, {Drimmel, R.},
  {H\o{}g, E.}, {Katz, D.}, {Lattanzi, M. G.}, {O\'{}Mullane, W.}, {Grebel, E.
  K.}, {Holland, A. D.}, {Huc, C.}, {Passot, X.}, {Bramante, L.}, {Cacciari,
  C.}, {Casta\~neda, J.}, {Chaoul, L.}, {Cheek, N.}, {De Angeli, F.},
  {Fabricius, C.}, {Guerra, R.}, {Hern\'andez, J.}, {Jean-Antoine-Piccolo, A.},
  {Masana, E.}, {Messineo, R.}, {Mowlavi, N.}, {Nienartowicz, K.},
  {Ord\'o\~nez-Blanco, D.}, {Panuzzo, P.}, {Portell, J.}, {Richards, P. J.},
  {Riello, M.}, {Seabroke, G. M.}, {Tanga, P.}, {Th\'evenin, F.}, {Torra, J.},
  {Els, S. G.}, {Gracia-Abril, G.}, {Comoretto, G.}, {Garcia-Reinaldos, M.},
  {Lock, T.}, {Mercier, E.}, {Altmann, M.}, {Andrae, R.}, {Astraatmadja, T.
  L.}, {Bellas-Velidis, I.}, {Benson, K.}, {Berthier, J.}, {Blomme, R.},
  {Busso, G.}, {Carry, B.}, {Cellino, A.}, {Clementini, G.}, {Cowell, S.},
  {Creevey, O.}, {Cuypers, J.}, {Davidson, M.}, {De Ridder, J.}, {de Torres,
  A.}, {Delchambre, L.}, {Dell\'{}Oro, A.}, {Ducourant, C.}, {Fr\'emat, Y.},
  {Garc\'{\i}a-Torres, M.}, {Gosset, E.}, {Halbwachs, J.-L.}, {Hambly, N. C.},
  {Harrison, D. L.}, {Hauser, M.}, {Hestroffer, D.}, {Hodgkin, S. T.}, {Huckle,
  H. E.}, {Hutton, A.}, {Jasniewicz, G.}, {Jordan, S.}, {Kontizas, M.}, {Korn,
  A. J.}, {Lanzafame, A. C.}, {Manteiga, M.}, {Moitinho, A.}, {Muinonen, K.},
  {Osinde, J.}, {Pancino, E.}, {Pauwels, T.}, {Petit, J.-M.}, {Recio-Blanco,
  A.}, {Robin, A. C.}, {Sarro, L. M.}, {Siopis, C.}, {Smith, M.}, {Smith, K.
  W.}, {Sozzetti, A.}, {Thuillot, W.}, {van Reeven, W.}, {Viala, Y.}, {Abbas,
  U.}, {Abreu Aramburu, A.}, {Accart, S.}, {Aguado, J. J.}, {Allan, P. M.},
  {Allasia, W.}, {Altavilla, G.}, {\'Alvarez, M. A.}, {Alves, J.}, {Anderson,
  R. I.}, {Andrei, A. H.}, {Anglada Varela, E.}, {Antiche, E.}, {Antoja, T.},
  {Ant\'on, S.}, {Arcay, B.}, {Atzei, A.}, {Ayache, L.}, {Bach, N.}, {Baker, S.
  G.}, {Balaguer-N\'u\~nez, L.}, {Barache, C.}, {Barata, C.}, {Barbier, A.},
  {Barblan, F.}, {Baroni, M.}, {Barrado y Navascu\'es, D.}, {Barros, M.},
  {Barstow, M. A.}, {Becciani, U.}, {Bellazzini, M.}, {Bellei, G.}, {Bello
  Garc\'{\i}a, A.}, {Belokurov, V.}, {Bendjoya, P.}, {Berihuete, A.}, {Bianchi,
  L.}, {Bienaym\'e, O.}, {Billebaud, F.}, {Blagorodnova, N.}, {Blanco-Cuaresma,
  S.}, {Boch, T.}, {Bombrun, A.}, {Borrachero, R.}, {Bouquillon, S.}, {Bourda,
  G.}, {Bouy, H.}, {Bragaglia, A.}, {Breddels, M. A.}, {Brouillet, N.},
  {Br\"usemeister, T.}, {Bucciarelli, B.}, {Budnik, F.}, {Burgess, P.},
  {Burgon, R.}, {Burlacu, A.}, {Busonero, D.}, {Buzzi, R.}, {Caffau, E.},
  {Cambras, J.}, {Campbell, H.}, {Cancelliere, R.}, {Cantat-Gaudin, T.},
  {Carlucci, T.}, {Carrasco, J. M.}, {Castellani, M.}, {Charlot, P.}, {Charnas,
  J.}, {Charvet, P.}, {Chassat, F.}, {Chiavassa, A.}, {Clotet, M.}, {Cocozza,
  G.}, {Collins, R. S.}, {Collins, P.}, {Costigan, G.}, {Crifo, F.}, {Cross, N.
  J. G.}, {Crosta, M.}, {Crowley, C.}, {Dafonte, C.}, {Damerdji, Y.},
  {Dapergolas, A.}, {David, P.}, {David, M.}, {De Cat, P.}, {de Felice, F.},
  {de Laverny, P.}, {De Luise, F.}, {De March, R.}, {de Martino, D.}, {de
  Souza, R.}, {Debosscher, J.}, {del Pozo, E.}, {Delbo, M.}, {Delgado, A.},
  {Delgado, H. E.}, {di Marco, F.}, {Di Matteo, P.}, {Diakite, S.}, {Distefano,
  E.}, {Dolding, C.}, {Dos Anjos, S.}, {Drazinos, P.}, {Dur\'an, J.}, {Dzigan,
  Y.}, {Ecale, E.}, {Edvardsson, B.}, {Enke, H.}, {Erdmann, M.}, {Escolar, D.},
  {Espina, M.}, {Evans, N. W.}, {Eynard Bontemps, G.}, {Fabre, C.}, {Fabrizio,
  M.}, {Faigler, S.}, {Falc\~ao, A. J.}, {Farr\`as Casas, M.}, {Faye, F.},
  {Federici, L.}, {Fedorets, G.}, {Fern\'andez-Hern\'andez, J.}, {Fernique,
  P.}, {Fienga, A.}, {Figueras, F.}, {Filippi, F.}, {Findeisen, K.}, {Fonti,
  A.}, {Fouesneau, M.}, {Fraile, E.}, {Fraser, M.}, {Fuchs, J.}, {Furnell, R.},
  {Gai, M.}, {Galleti, S.}, {Galluccio, L.}, {Garabato, D.},
  {Garc\'{\i}a-Sedano, F.}, {Gar\'e, P.}, {Garofalo, A.}, {Garralda, N.},
  {Gavras, P.}, {Gerssen, J.}, {Geyer, R.}, {Gilmore, G.}, {Girona, S.},
  {Giuffrida, G.}, {Gomes, M.}, {Gonz\'alez-Marcos, A.}, {Gonz\'alez-N\'u\~nez,
  J.}, {Gonz\'alez-Vidal, J. J.}, {Granvik, M.}, {Guerrier, A.}, {Guillout,
  P.}, {Guiraud, J.}, {G\'urpide, A.}, {Guti\'errez-S\'anchez, R.}, {Guy, L.
  P.}, {Haigron, R.}, {Hatzidimitriou, D.}, {Haywood, M.}, {Heiter, U.},
  {Helmi, A.}, {Hobbs, D.}, {Hofmann, W.}, {Holl, B.}, {Holland, G.}, {Hunt, J.
  A. S.}, {Hypki, A.}, {Icardi, V.}, {Irwin, M.}, {Jevardat de Fombelle, G.},
  {Jofr\'e, P.}, {Jonker, P. G.}, {Jorissen, A.}, {Julbe, F.}, {Karampelas,
  A.}, {Kochoska, A.}, {Kohley, R.}, {Kolenberg, K.}, {Kontizas, E.}, {Koposov,
  S. E.}, {Kordopatis, G.}, {Koubsky, P.}, {Kowalczyk, A.}, {Krone-Martins,
  A.}, {Kudryashova, M.}, {Kull, I.}, {Bachchan, R. K.}, {Lacoste-Seris, F.},
  {Lanza, A. F.}, {Lavigne, J.-B.}, {Le Poncin-Lafitte, C.}, {Lebreton, Y.},
  {Lebzelter, T.}, {Leccia, S.}, {Leclerc, N.}, {Lecoeur-Taibi, I.}, {Lemaitre,
  V.}, {Lenhardt, H.}, {Leroux, F.}, {Liao, S.}, {Licata, E.}, {Lindstr\o{}m,
  H. E. P.}, {Lister, T. A.}, {Livanou, E.}, {Lobel, A.}, {L\"offler, W.},
  {L\'opez, M.}, {Lopez-Lozano, A.}, {Lorenz, D.}, {Loureiro, T.}, {MacDonald,
  I.}, {Magalh\~aes Fernandes, T.}, {Managau, S.}, {Mann, R. G.}, {Mantelet,
  G.}, {Marchal, O.}, {Marchant, J. M.}, {Marconi, M.}, {Marie, J.}, {Marinoni,
  S.}, {Marrese, P. M.}, {Marschalk\'o, G.}, {Marshall, D. J.},
  {Mart\'{\i}n-Fleitas, J. M.}, {Martino, M.}, {Mary, N.}, {Matijevic, G.},
  {Mazeh, T.}, {McMillan, P. J.}, {Messina, S.}, {Mestre, A.}, {Michalik, D.},
  {Millar, N. R.}, {Miranda, B. M. H.}, {Molina, D.}, {Molinaro, R.},
  {Molinaro, M.}, {Moln\'ar, L.}, {Moniez, M.}, {Montegriffo, P.}, {Monteiro,
  D.}, {Mor, R.}, {Mora, A.}, {Morbidelli, R.}, {Morel, T.}, {Morgenthaler,
  S.}, {Morley, T.}, {Morris, D.}, {Mulone, A. F.}, {Muraveva, T.}, {Musella,
  I.}, {Narbonne, J.}, {Nelemans, G.}, {Nicastro, L.}, {Noval, L.},
  {Ord\'enovic, C.}, {Ordieres-Mer\'e, J.}, {Osborne, P.}, {Pagani, C.},
  {Pagano, I.}, {Pailler, F.}, {Palacin, H.}, {Palaversa, L.}, {Parsons, P.},
  {Paulsen, T.}, {Pecoraro, M.}, {Pedrosa, R.}, {Pentik\"ainen, H.}, {Pereira,
  J.}, {Pichon, B.}, {Piersimoni, A. M.}, {Pineau, F.-X.}, {Plachy, E.}, {Plum,
  G.}, {Poujoulet, E.}, {Prsa, A.}, {Pulone, L.}, {Ragaini, S.}, {Rago, S.},
  {Rambaux, N.}, {Ramos-Lerate, M.}, {Ranalli, P.}, {Rauw, G.}, {Read, A.},
  {Regibo, S.}, {Renk, F.}, {Reyl\'e, C.}, {Ribeiro, R. A.}, {Rimoldini, L.},
  {Ripepi, V.}, {Riva, A.}, {Rixon, G.}, {Roelens, M.}, {Romero-G\'omez, M.},
  {Rowell, N.}, {Royer, F.}, {Rudolph, A.}, {Ruiz-Dern, L.}, {Sadowski, G.},
  {Sagrist\`a Sell\'es, T.}, {Sahlmann, J.}, {Salgado, J.}, {Salguero, E.},
  {Sarasso, M.}, {Savietto, H.}, {Schnorhk, A.}, {Schultheis, M.}, {Sciacca,
  E.}, {Segol, M.}, {Segovia, J. C.}, {Segransan, D.}, {Serpell, E.}, {Shih,
  I-C.}, {Smareglia, R.}, {Smart, R. L.}, {Smith, C.}, {Solano, E.}, {Solitro,
  F.}, {Sordo, R.}, {Soria Nieto, S.}, {Souchay, J.}, {Spagna, A.}, {Spoto,
  F.}, {Stampa, U.}, {Steele, I. A.}, {Steidelm\"uller, H.}, {Stephenson, C.
  A.}, {Stoev, H.}, {Suess, F. F.}, {S\"uveges, M.}, {Surdej, J.}, {Szabados,
  L.}, {Szegedi-Elek, E.}, {Tapiador, D.}, {Taris, F.}, {Tauran, G.}, {Taylor,
  M. B.}, {Teixeira, R.}, {Terrett, D.}, {Tingley, B.}, {Trager, S. C.},
  {Turon, C.}, {Ulla, A.}, {Utrilla, E.}, {Valentini, G.}, {van Elteren, A.},
  {Van Hemelryck, E.}, {van Leeuwen, M.}, {Varadi, M.}, {Vecchiato, A.},
  {Veljanoski, J.}, {Via, T.}, {Vicente, D.}, {Vogt, S.}, {Voss, H.}, {Votruba,
  V.}, {Voutsinas, S.}, {Walmsley, G.}, {Weiler, M.}, {Weingrill, K.}, {Werner,
  D.}, {Wevers, T.}, {Whitehead, G.}, {Wyrzykowski, L.}, {Yoldas, A.}, {Zerjal,
  M.}, {Zucker, S.}, {Zurbach, C.}, {Zwitter, T.}, {Alecu, A.}, {Allen, M.},
  {Allende Prieto, C.}, {Amorim, A.}, {Anglada-Escud\'e, G.}, {Arsenijevic,
  V.}, {Azaz, S.}, {Balm, P.}, {Beck, M.}, {Bernstein, H.-H.}, {Bigot, L.},
  {Bijaoui, A.}, {Blasco, C.}, {Bonfigli, M.}, {Bono, G.}, {Boudreault, S.},
  {Bressan, A.}, {Brown, S.}, {Brunet, P.-M.}, {Bunclark, P.}, {Buonanno, R.},
  {Butkevich, A. G.}, {Carret, C.}, {Carrion, C.}, {Chemin, L.}, {Ch\'ereau,
  F.}, {Corcione, L.}, {Darmigny, E.}, {de Boer, K. S.}, {de Teodoro, P.}, {de
  Zeeuw, P. T.}, {Delle Luche, C.}, {Domingues, C. D.}, {Dubath, P.}, {Fodor,
  F.}, {Fr\'ezouls, B.}, {Fries, A.}, {Fustes, D.}, {Fyfe, D.}, {Gallardo, E.},
  {Gallegos, J.}, {Gardiol, D.}, {Gebran, M.}, {Gomboc, A.}, {G\'omez, A.},
  {Grux, E.}, {Gueguen, A.}, {Heyrovsky, A.}, {Hoar, J.}, {Iannicola, G.},
  {Isasi Parache, Y.}, {Janotto, A.-M.}, {Joliet, E.}, {Jonckheere, A.}, {Keil,
  R.}, {Kim, D.-W.}, {Klagyivik, P.}, {Klar, J.}, {Knude, J.}, {Kochukhov, O.},
  {Kolka, I.}, {Kos, J.}, {Kutka, A.}, {Lainey, V.}, {LeBouquin, D.}, {Liu,
  C.}, {Loreggia, D.}, {Makarov, V. V.}, {Marseille, M. G.}, {Martayan, C.},
  {Martinez-Rubi, O.}, {Massart, B.}, {Meynadier, F.}, {Mignot, S.}, {Munari,
  U.}, {Nguyen, A.-T.}, {Nordlander, T.}, {Ocvirk, P.}, {O\'{}Flaherty, K. S.},
  {Olias Sanz, A.}, {Ortiz, P.}, {Osorio, J.}, {Oszkiewicz, D.}, {Ouzounis,
  A.}, {Palmer, M.}, {Park, P.}, {Pasquato, E.}, {Peltzer, C.}, {Peralta, J.},
  {P\'eturaud, F.}, {Pieniluoma, T.}, {Pigozzi, E.}, {Poels, J.}, {Prat, G.},
  {Prod\'{}homme, T.}, {Raison, F.}, {Rebordao, J. M.}, {Risquez, D.},
  {Rocca-Volmerange, B.}, {Rosen, S.}, {Ruiz-Fuertes, M. I.}, {Russo, F.},
  {Sembay, S.}, {Serraller Vizcaino, I.}, {Short, A.}, {Siebert, A.}, {Silva,
  H.}, {Sinachopoulos, D.}, {Slezak, E.}, {Soffel, M.}, {Sosnowska, D.},
  {Straizys, V.}, {ter Linden, M.}, {Terrell, D.}, {Theil, S.}, {Tiede, C.},
  {Troisi, L.}, {Tsalmantza, P.}, {Tur, D.}, {Vaccari, M.}, {Vachier, F.},
  {Valles, P.}, {Van Hamme, W.}, {Veltz, L.}, {Virtanen, J.}, {Wallut, J.-M.},
  {Wichmann, R.}, {Wilkinson, M. I.}, {Ziaeepour, H.}, \& {Zschocke,
  S.}}]{gaia2016b}
{Gaia Collaboration}, {Prusti, T.}, {de Bruijne, J. H. J.}, {et~al.} 2016,
  A\&A, 595, A1

\bibitem[{{Gaia Collaboration} {et~al.}(2022){Gaia Collaboration}, Vallenari,
  Brown, Prusti, de~Bruijne, Arenou, Babusiaux, Biermann, Creevey, Ducourant,
  Evans, Eyer, Guerra, Hutton, Jordi, Klioner, Lammers, Lindegren, Luri,
  Mignard, Panem, Pourbaix, Randich, Sartoretti, Soubiran, Tanga, Walton,
  Bailer-Jones, Bastian, Drimmel, Jansen, Katz, Lattanzi, van Leeuwen, Bakker,
  Cacciari, Castañeda, Angeli, Fabricius, Fouesneau, Frémat, Galluccio,
  Guerrier, Heiter, Masana, Messineo, Mowlavi, Nicolas, Nienartowicz, Pailler,
  Panuzzo, Riclet, Roux, Seabroke, Sordoørcit, Thévenin, Gracia-Abril,
  Portell, Teyssier, Altmann, Andrae, Audard, Bellas-Velidis, Benson, Berthier,
  Blomme, Burgess, Busonero, Busso, Cánovas, Carry, Cellino, Cheek,
  Clementini, Damerdji, Davidson, de~Teodoro, Campos, Delchambre, Dell'Oro,
  Esquej, Fernández-Hernández, Fraile, Garabato, García-Lario, Gosset,
  Haigron, Halbwachs, Hambly, Harrison, Hernández, Hestroffer, Hodgkin, Holl,
  Janßen, de~Fombelle, Jordan, Krone-Martins, Lanzafame, Löffler, Marchal,
  Marrese, Moitinho, Muinonen, Osborne, Pancino, Pauwels, Recio-Blanco, Reylé,
  Riello, Rimoldini, Roegiers, Rybizki, Sarro, Siopis, Smith, Sozzetti,
  Utrilla, van Leeuwen, Abbas, Ábrahám, Aramburu, Aerts, Aguado, Ajaj,
  Aldea-Montero, Altavilla, Álvarez, Alves, Anders, Anderson, Varela, Antoja,
  Baines, Baker, Balaguer-Núñez, Balbinot, Balog, Barache, Barbato, Barros,
  Barstow, Bartolomé, Bassilana, Bauchet, Becciani, Bellazzini, Berihuete,
  Bernet, Bertone, Bianchi, Binnenfeld, Blanco-Cuaresma, Blazere, Boch,
  Bombrun, Bossini, Bouquillon, Bragaglia, Bramante, Breedt, Bressan,
  Brouillet, Brugaletta, Bucciarelli, Burlacu, Butkevich, Buzzi, Caffau,
  Cancelliere, Cantat-Gaudin, Carballo, Carlucci, Carnerero, Carrasco,
  Casamiquela, Castellani, Castro-Ginard, Chaoul, Charlot, Chemin, Chiaramida,
  Chiavassa, Chornay, Comoretto, Contursi, Cooper, Cornez, Cowell, Crifo,
  Cropper, Crosta, Crowley, Dafonte, Dapergolas, David, David, de~Laverny,
  Luise, March, Ridder, de~Souza, de~Torres, del Peloso, del Pozo, Delbo,
  Delgado, Delisle, Demouchy, Dharmawardena, Matteo, Diakite, Diener,
  Distefano, Dolding, Edvardsson, Enke, Fabre, Fabrizio, Faigler, Fedorets,
  Fernique, Fienga, Figueras, Fournier, Fouron, Fragkoudi, Gai,
  Garcia-Gutierrez, Garcia-Reinaldos, García-Torres, Garofalo, Gavel, Gavras,
  Gerlach, Geyer, Giacobbe, Gilmore, Girona, Giuffrida, Gomel, Gomez,
  González-Núñez, González-Santamaría, González-Vidal, Granvik, Guillout,
  Guiraud, Gutiérrez-Sánchez, Guy, Hatzidimitriou, Hauser, Haywood, Helmer,
  Helmi, Sarmiento, Hidalgo, Hilger, Hładczuk, Hobbs, Holland, Huckle,
  Jardine, Jasniewicz, Piccolo, Jiménez-Arranz, Jorissen, Campillo, Julbe,
  Karbevska, Kervella, Khanna, Kontizas, Kordopatis, Korn, Kóspál,
  Kostrzewa-Rutkowska, Kruszyńska, Kun, Laizeau, Lambert, Lanza, Lasne,
  Campion, Lebreton, Lebzelter, Leccia, Leclerc, Lecoeur-Taibi, Liao, Licata,
  Lindstrøm, Lister, Livanou, Lobel, Lorca, Loup, Pardo, Romeo, Managau, Mann,
  Manteiga, Marchant, Marconi, Marcos, Santos, Pina, Marinoni, Marocco,
  Marshall, Polo, Martín-Fleitas, Marton, Mary, Masip, Massari,
  Mastrobuono-Battisti, Mazeh, McMillan, Messina, Michalik, Millar, Mints,
  Molina, Molinaro, Molnár, Monari, Monguió, Montegriffo, Montero, Mor, Mora,
  Morbidelli, Morel, Morris, Muraveva, Murphy, Musella, Nagy, Noval, Ocaña,
  Ogden, Ordenovic, Osinde, Pagani, Pagano, Palaversa, Palicio,
  Pallas-Quintela, Panahi, Payne-Wardenaar, Esteller, Penttilä, Pichon,
  Piersimoni, Pineau, Plachy, Plum, Poggio, Prša, Pulone, Racero, Ragaini,
  Rainer, Raiteri, Rambaux, Ramos, Ramos-Lerate, Fiorentin, Regibo, Richards,
  Diaz, Ripepi, Riva, Rix, Rixon, Robichon, Robin, Robin, Roelens, Rogues,
  Rohrbasser, Romero-Gómez, Rowell, Royer, Mieres, Rybicki, Sadowski, Núñez,
  Sellés, Sahlmann, Salguero, Samaras, Gimenez, Sanna, Santoveña, Sarasso,
  Schultheis, Sciacca, Segol, Segovia, Ségransan, Semeux, Shahaf, Siddiqui,
  Siebert, Siltala, Silvelo, Slezak, Slezak, Smart, Snaith, Solano, Solitro,
  Souami, Souchay, Spagna, Spina, Spoto, Steele, Steidelmüller, Stephenson,
  Süveges, Surdej, Szabados, Szegedi-Elek, Taris, Taylo, Teixeira, Tolomei,
  Tonello, Torra, Torra, Elipe, Trabucchi, Tsounis, Turon, Ulla, Unger,
  Vaillant, van Dillen, van Reeven, Vanel, Vecchiato, Viala, Vicente,
  Voutsinas, Weiler, Wevers, Wyrzykowski, Yoldas, Yvard, Zhao, Zorec, Zucker,
  \& Zwitter}]{gaia2022k}
{Gaia Collaboration}, Vallenari, A., Brown, A. G.~A., {et~al.} 2022, Gaia Data
  Release 3: Summary of the content and survey properties, , , arXiv:2208.00211

\bibitem[{Gray \& Corbally(1994)}]{gray94}
Gray, R.~O., \& Corbally, C.~J. 1994, AJ, 107, 742

\bibitem[{Gray \& Corbally(2009)}]{gray09}
---. 2009, Stellar Spectral Classification (Princeton, New Jersey: Princeton
  University Press)

\bibitem[{Gray \& Corbally(2014)}]{gray14}
---. 2014, AJ, 147, 80

\bibitem[{Gray \& Garrison(1987)}]{gray87}
Gray, R.~O., \& Garrison, R.~F. 1987, ApJS, 65, 581

\bibitem[{Green {et~al.}(2019)Green, Schlafly, Zucker, Seagle, \&
  Finkbeiner}]{green19}
Green, G.~M., Schlafly, E., Zucker, C., Seagle, J.~S., \& Finkbeiner, D. 2019,
  ApJ, 887, 93

\bibitem[{{Henden} {et~al.}(2009){Henden}, {Welch}, {Terrell}, \&
  {Levine}}]{henden09}
{Henden}, A.~A., {Welch}, D.~L., {Terrell}, D., \& {Levine}, S.~E. 2009, in
  American Astronomical Society Meeting Abstracts, Vol. 214, American
  Astronomical Society Meeting Abstracts \#214, 407.02

\bibitem[{Horne(1986)}]{horne86}
Horne, K. 1986, PASP, 98

\bibitem[{Kervella {et~al.}(2019)Kervella, Arenou, Mignard, \&
  Thevenin}]{kervella19}
Kervella, P., Arenou, F., Mignard, F., \& Thevenin, F. 2019, A\&A, 623, A72: 1

\bibitem[{Koen \& Eyer(2002)}]{koen02}
Koen, C., \& Eyer, L. 2002, MNRAS, 331, 45

\bibitem[{Mason {et~al.}(2001)Mason, Hartkopf, Holdenried, \&
  Rafferty}]{mason01}
Mason, B.~D., Hartkopf, W.~I., Holdenried, E.~R., \& Rafferty, T.~J. 2001, AJ,
  121, 3224

\bibitem[{McClure {et~al.}(1980)McClure, Fletcher, \& Nemec}]{mcclure80}
McClure, R.~D., Fletcher, J.~M., \& Nemec, J.~M. 1980, ApJ, 238, L35

\bibitem[{North(2000)}]{north00}
North, P. 2000, in International Astronomical Union Symposia, Vol. 177, The
  Carbon Star Phenomenon, Proceedings of the 177th Symposium of the
  International Astronomical Union, held in Antalya, Turkey, May 27-31, 1996,
  ed. R.~F. Wing (Kluwer Academic Publishers, Dordrecht), 269--275

\bibitem[{Perry \& Johnston(1982)}]{perry82}
Perry, C.~L., \& Johnston, L. 1982, ApJSS, 50, 451

\bibitem[{Reiter {et~al.}(2015)Reiter, Marengo, Hora, \& Fazio}]{reiter15}
Reiter, M., Marengo, M., Hora, J.~L., \& Fazio, G.~G. 2015, MNRAS, 447, 3909

\bibitem[{Richer(1972)}]{richer72}
Richer, H. 1972, ApJ, 172, L63

\bibitem[{Samus {et~al.}(2017)Samus, Kazarovets, Durlevich, Kireeva, \&
  Pastukhova}]{gcvs}
Samus, N.~N., Kazarovets, E.~V., Durlevich, O.~N., Kireeva, N.~N., \&
  Pastukhova, E.~N. 2017, Astronomy Reports, 61, 80

\bibitem[{Sanford(1944)}]{sanford44}
Sanford, R.~F. 1944, ApJ, 99, 145

\bibitem[{{The Astropy Collaboration} {et~al.}(2013){The Astropy
  Collaboration}, {Robitaille, Thomas P.}, {Tollerud, Erik J.}, {Greenfield,
  Perry}, {Droettboom, Michael}, {Bray, Erik}, {Aldcroft, Tom}, {Davis, Matt},
  {Ginsburg, Adam}, {Price-Whelan, Adrian M.}, {Kerzendorf, Wolfgang E.},
  {Conley, Alexander}, {Crighton, Neil}, {Barbary, Kyle}, {Muna, Demitri},
  {Ferguson, Henry}, {Grollier, Fr\'ed\'eric}, {Parikh, Madhura M.}, {Nair,
  Prasanth H.}, {G\"unther, Hans M.}, {Deil, Christoph}, {Woillez, Julien},
  {Conseil, Simon}, {Kramer, Roban}, {Turner, James E. H.}, {Singer, Leo},
  {Fox, Ryan}, {Weaver, Benjamin A.}, {Zabalza, Victor}, {Edwards, Zachary I.},
  {Azalee Bostroem, K.}, {Burke, D. J.}, {Casey, Andrew R.}, {Crawford, Steven
  M.}, {Dencheva, Nadia}, {Ely, Justin}, {Jenness, Tim}, {Labrie, Kathleen},
  {Lim, Pey Lian}, {Pierfederici, Francesco}, {Pontzen, Andrew}, {Ptak, Andy},
  {Refsdal, Brian}, {Servillat, Mathieu}, \& {Streicher, Ole}}]{astropy2013}
{The Astropy Collaboration}, {Robitaille, Thomas P.}, {Tollerud, Erik J.},
  {et~al.} 2013, A\&A, 558, A33

\bibitem[{{The Astropy Collaboration} {et~al.}(2018){The Astropy
  Collaboration}, Price-Whelan, Sip{\H{o} }cz, Günther, Lim, Crawford,
  Conseil, Shupe, Craig, Dencheva, Ginsburg, VanderPlas, Bradley,
  P{\'{e}}rez-Su{\'{a}}rez, de~Val-Borro, Aldcroft, Cruz, Robitaille, Tollerud,
  Ardelean, Babej, Bach, Bachetti, Bakanov, Bamford, Barentsen, Barmby,
  Baumbach, Berry, Biscani, Boquien, Bostroem, Bouma, Brammer, Bray,
  Breytenbach, Buddelmeijer, Burke, Calderone, Rodr{\'{\i}}guez, Cara, Cardoso,
  Cheedella, Copin, Corrales, Crichton, D'Avella, Deil, Depagne, Dietrich,
  Donath, Droettboom, Earl, Erben, Fabbro, Ferreira, Finethy, Fox, Garrison,
  Gibbons, Goldstein, Gommers, Greco, Greenfield, Groener, Grollier, Hagen,
  Hirst, Homeier, Horton, Hosseinzadeh, Hu, Hunkeler, Ivezi{\'{c}}, Jain,
  Jenness, Kanarek, Kendrew, Kern, Kerzendorf, Khvalko, King, Kirkby, Kulkarni,
  Kumar, Lee, Lenz, Littlefair, Ma, Macleod, Mastropietro, McCully, Montagnac,
  Morris, Mueller, Mumford, Muna, Murphy, Nelson, Nguyen, Ninan, Nöthe, Ogaz,
  Oh, Parejko, Parley, Pascual, Patil, Patil, Plunkett, Prochaska, Rastogi,
  Janga, Sabater, Sakurikar, Seifert, Sherbert, Sherwood-Taylor, Shih, Sick,
  Silbiger, Singanamalla, Singer, Sladen, Sooley, Sornarajah, Streicher,
  Teuben, Thomas, Tremblay, Turner, Terr{\'{o}}n, van Kerkwijk, de~la Vega,
  Watkins, Weaver, Whitmore, Woillez, \& Zabalza}]{astropy2018}
{The Astropy Collaboration}, Price-Whelan, A.~M., Sip{\H{o} }cz, B.~M.,
  {et~al.} 2018, AJ, 156, 123

\bibitem[{{The Astropy Collaboration} {et~al.}(2022){The Astropy
  Collaboration}, {Price-Whelan}, {Lim}, {Earl}, {Starkman}, {Bradley},
  {Shupe}, {Patil}, {Corrales}, {Brasseur}, {N{\"o}the}, {Donath}, {Tollerud},
  {Morris}, {Ginsburg}, {Vaher}, {Weaver}, {Tocknell}, {Jamieson}, {van
  Kerkwijk}, {Robitaille}, {Merry}, {Bachetti}, {G{\"u}nther}, {Aldcroft},
  {Alvarado-Montes}, {Archibald}, {B{\'o}di}, {Bapat}, {Barentsen},
  {Baz{\'a}n}, {Biswas}, {Boquien}, {Burke}, {Cara}, {Cara}, {Conroy},
  {Conseil}, {Craig}, {Cross}, {Cruz}, {D'Eugenio}, {Dencheva}, {Devillepoix},
  {Dietrich}, {Eigenbrot}, {Erben}, {Ferreira}, {Foreman-Mackey}, {Fox},
  {Freij}, {Garg}, {Geda}, {Glattly}, {Gondhalekar}, {Gordon}, {Grant},
  {Greenfield}, {Groener}, {Guest}, {Gurovich}, {Handberg}, {Hart},
  {Hatfield-Dodds}, {Homeier}, {Hosseinzadeh}, {Jenness}, {Jones}, {Joseph},
  {Kalmbach}, {Karamehmetoglu}, {Ka{\l}uszy{\'n}ski}, {Kelley}, {Kern},
  {Kerzendorf}, {Koch}, {Kulumani}, {Lee}, {Ly}, {Ma}, {MacBride}, {Maljaars},
  {Muna}, {Murphy}, {Norman}, {O'Steen}, {Oman}, {Pacifici}, {Pascual},
  {Pascual-Granado}, {Patil}, {Perren}, {Pickering}, {Rastogi}, {Roulston},
  {Ryan}, {Rykoff}, {Sabater}, {Sakurikar}, {Salgado}, {Sanghi}, {Saunders},
  {Savchenko}, {Schwardt}, {Seifert-Eckert}, {Shih}, {Jain}, {Shukla}, {Sick},
  {Simpson}, {Singanamalla}, {Singer}, {Singhal}, {Sinha}, {Sip{\H{o}}cz},
  {Spitler}, {Stansby}, {Streicher}, {{\v{S}}umak}, {Swinbank}, {Taranu},
  {Tewary}, {Tremblay}, {de Val-Borro}, {Van Kooten}, {Vasovi{\'c}}, {Verma},
  {de Miranda Cardoso}, {Williams}, {Wilson}, {Winkel}, {Wood-Vasey}, {Xue},
  {Yoachim}, {Zhang}, {Zonca}, \& {Astropy Project Contributors}}]{astropy2022}
{The Astropy Collaboration}, {Price-Whelan}, A.~M., {Lim}, P.~L., {et~al.}
  2022, \apj, 935, 167

\bibitem[{Tody(1986)}]{iraf86}
Tody, D. 1986, in Society of Photo-Optical Instrumentation Engineers (SPIE)
  Conference Series, Vol. 627, Instrumentation in astronomy VI, ed. D.~L.
  {Crawford}, 733

\bibitem[{Tody(1993)}]{iraf93}
Tody, D. 1993, in Astronomical Society of the Pacific Conference Series,
  Vol.~52, Astronomical Data Analysis Software and Systems II, ed. R.~J.
  {Hanisch}, R.~J.~V. {Brissenden}, \& J.~{Barnes}, 173

\bibitem[{van Belle {et~al.}(2013)van Belle, Paladini, Aringer, Hron, \&
  Ciardi}]{vanbelle13}
van Belle, G.~T., Paladini, C., Aringer, B., Hron, J., \& Ciardi, D. 2013, ApJ,
  775, 45

\bibitem[{Wang {et~al.}(1996)Wang, Su, Chu, Cui, \& Wang}]{wang96}
Wang, S.-G., Su, D.-Q., Chu, Y.-Q., Cui, X., \& Wang, Y. 1996, ApOpt, 35, 5155

\end{thebibliography}
\bibliographystyle{aasjournal}

\end{document}